\documentclass[a4paper,twocolumn,11pt,accepted=2022-08-24]{quantumarticle}
\pdfoutput=1
\usepackage[utf8]{inputenc}
\usepackage[english]{babel}
\usepackage[T1]{fontenc}

\usepackage{amsmath}
\usepackage{amssymb}
\usepackage{hyperref}

\usepackage{graphicx}

\title{Critical quantum thermometry and its feasibility in spin systems}

\author{Enes Aybar}
\affiliation{ICFO - Institut de Ciencies Fotoniques, The Barcelona Institute of Science and Technology, 08860 Castelldefels, Barcelona, Spain}
\orcid{0000-0001-6946-4098}

\author{Artur Niezgoda}
\affiliation{ICFO - Institut de Ciencies Fotoniques, The Barcelona Institute of Science and Technology, 08860 Castelldefels, Barcelona, Spain}
\affiliation{Faculty of Physics, University of Warsaw, ul. Pasteura 5, PL-02-093 Warsaw, Poland}
\orcid{0000-0002-2263-2725}
\author{Safoura S. Mirkhalaf}
\affiliation{Department of Physics, University of Tehran, P.O. Box 14395-547, Tehran, Iran}
\affiliation{School of Nano Science, Institute for Research in Fundamental Sciences (IPM), P.O. Box 19395-5531, Tehran, Iran}
\orcid{0000-0002-4154-8465}

\author{Morgan W. Mitchell}
\affiliation{ICFO - Institut de Ciencies Fotoniques, The Barcelona Institute of Science and Technology, 08860 Castelldefels, Barcelona, Spain}
\affiliation{ICREA - Instituci\'{o} Catalana de Recerca i Estudis Avan{\c{c}}ats, 08010 Barcelona, Spain}
\orcid{0000-0001-8949-9407}

\author{Daniel Benedicto Orenes}
\affiliation{ICFO - Institut de Ciencies Fotoniques, The Barcelona Institute of Science and Technology, 08860 Castelldefels, Barcelona, Spain}
\orcid{0000-0002-8098-012X}

\author{Emilia Witkowska}
\affiliation{Institute of Physics PAS, Aleja Lotnikow 32/46, 02-668 Warszawa, Poland}
\orcid{0000-0003-4622-8513}

\begin{document}
\maketitle

\begin{abstract}
In this work, we study temperature sensing with finite-sized strongly correlated systems exhibiting quantum phase transitions. We use the quantum Fisher information (QFI) approach to quantify the sensitivity in the temperature estimation, and apply a finite-size scaling framework to link this sensitivity to critical exponents of the system around critical points. We numerically calculate the QFI around the critical points for two experimentally-realizable systems: the spin-1 Bose-Einstein condensate and the spin-chain Heisenberg XX model in the presence of an external magnetic field. Our results confirm finite-size scaling properties of the QFI. Furthermore, we discuss experimentally-accessible observables that (nearly) saturate the QFI at the critical points for these two systems. 
\end{abstract}

\section{Introduction}

Understanding and quantifying uncertainty in the measurement of a classical parameter using quantum resources is a central goal of modern estimation theory~\cite{Helstrom1969} with far-reaching implications both for fundamental studies and quantum technology applications~\cite{GobelSiegner2015}. In the well-studied scenario of single-parameter estimation, the geometry of quantum states~\cite{Braunstein1994}, connects the sensitivity of parameter estimation to the fidelity between neighboring states.  Concretely, if $\hat{\rho}(\lambda)$ is a parametrized state and $\mathcal{F}_{\lambda,\lambda'}$ is the fidelity between $\hat{\rho}(\lambda)$ and $\hat{\rho}(\lambda')$, the quantum Fisher information (QFI) can be related to the second derivative $\partial^2_{\lambda'} \mathcal{F}_{\lambda,\lambda'}|_{\lambda'=\lambda}$~\cite{Braunstein1994, Taddei2013, Toth2014}; when this quantity is large, $\lambda$ can in principle be estimated with high precision.

In order to enhance this sensitivity different resources have been studied, including: entanglement~\cite{Pezze2014}, inter-particle interaction~\cite{Napolitano2011} and quantum phase transitions (QPTs)~\cite{Zanardi2008}. Regarding the latter, prior work has studied parametrized states, e.g., ground states of a parametrized Hamiltonian $\hat{H}(\lambda)$. The rapid state changes at the QPT imply larger QFI for $\lambda$, i.e., for  the QPT's control parameter~\cite{Zanardi2008, Mok2021, Gietka2021, Chu2021, Garbe2020, Rams2018, Mirkhalaf2020, Mirkhalaf2021, Pezze2019, Salvatori2014, Tsang2013}. A relation between sensitivity to $\lambda$ and universal scaling behaviour near critical points has been established ~\cite{Zanardi2007_1, You2007, Hauke2016, Gu2010}.

Thermometry is another paradigmatic example of a quantum estimation problem; temperature $T$ is not an observable in the quantum mechanical sense, but rather a parameter that can be estimated from measurements of proper observables. The study of fundamental limits to sensitivity in temperature estimation has attracted attention in different areas: from thermalization studies of open quantum systems~\cite{Ashida2018} and characterization of trapped ions for quantum computation~\cite{Ivanov2019} to temperature sensing by micro-organisms~\cite{Vennettilli2021}.

Here we study thermometry in quantum critical systems using the tools of quantum parameter estimation. The parametrized state $\hat{\rho}(T,\lambda)$ of interest is the Gibbs state of a parametrized Hamiltonian $\hat{H}(\lambda)$. We focus on systems that exhibit continuous QPTs at zero temperature and in the thermodynamic limit. We explore their behaviour for finite numbers of particles $N$ and finite temperature $T$. In addition, we find relations between the sensitivity of temperature estimation and  universal scaling behaviours  around the critical points. At zero temperature and in the thermodynamic limit, a QPT shows power-law divergences of physical  quantities, described by critical exponents that quantify the speed of the divergence~\cite{Continentino2001}. When the system size is finite, the behaviour of physical quantities are regular at critical points and can be described by analytic functions that are subject to a universal finite-size scaling~\cite{Continentino2001, Cardy1988, Campostrini2014}. At small but finite temperatures, Gibbs states can inherit signatures of quantum critical behavior that, as in the zero-temperature case,  enhance sensitivity in parameter estimation around the QPT~\cite{Zanardi2007_2, Hauke2016, Zanardi2007_3}.

We quantify the sensitivity in temperature estimation using the quantum Fisher information and relate it with the critical exponents at the QPT through a universal scaling function that is directly proportional to the inverse of the lowest energy gap squared.   This relationship causes the abrupt growth of the QFI in the critical region when temperature is lower than the energy gap. In addition, we analyze the signal-to-noise ratio (SNR) showing its relation with critical exponents and the associated scaling function. We find out that the SNR does not exhibit the direct dependence on energy gap, i.e., diminishing the energy gap in the critical region does not lead to a larger SNR.

Finally, we illustrate the resulting finite-size scaling properties of the sensitivity using two interesting and experimentally-relevant examples: (i) the ferromagnetic spin-1 Bose-Einstein condensate (BEC) and (ii) the ferromagnetic XX spin-1/2 chain, both subject to the presence of an external magnetic field. The two systems considered by us can be realized with current experimental techniques using ultra-cold atoms~\cite{Zou2018, Jepsen2020}. We discuss possible experimental implementations, calculating the sensitivity of temperature measurements in these systems for different experimentally-accessible observables identifying the ones that would  saturate the QFI. Our results demonstrate the validity of the finite-size scaling analysis and show how quantum thermometry in strongly correlated systems exhibiting continuous QPTs can benefit from the criticality when comparing to their far-from-criticality behavior. In this sense we refer to \textit{criticality-enhanced quantum thermometry}. This enhancement in the sensitivity can lead to the realization of useful thermometers based on spin-measurements, i.e., not relying on external ancillary probes~\cite{Hohmann2016, Bouton2020}, further manipulation of the atomic sample~\cite{Leanhardt2003} or thermal fraction of the ultra-cold quantum gases~\cite{Olf2015}.

\section{Local Quantum Thermometry and Criticality}\label{sec:theoryofquantumthermometry}
In this section we present the principles of temperature estimation and its sensitivity bounds quantified by the QFI. We introduce the concept of a finite-size scaling framework for systems exhibiting continuous QPTs. The scaling approach is then applied to study the behavior of the QFI in the vicinity of QPTs.

\subsection{QFI approach for quantum thermometry}
Thermometry is the classical parameter estimation problem conveniently resolved using the parameter estimation protocols~\cite{Paris2015, Mehboudi2019_1}. More precisely, the temperature of a system consisting of $N$ particles is inferred from results of $m$ measurements $\{\alpha_i\}$ of an observable $\hat{A}$ of the system under consideration through the estimator function $ T_{\rm est} (\{\alpha_i \})$ which we assume to be unbiased, i.e., $\langle T_{\rm est}\rangle = T$ ($i$ labels independent measurements). We define the resulting fluctuations of temperature in terms of the mean squared error of the corresponding estimator $\delta^2T \equiv  \langle ( T_{\rm est} -  T)^2 \rangle$ which is subject to the Cram\'er-Rao lower bound~\cite{Cramer1999}:
\begin{equation}\label{eq:CRLB}
    \delta^2  T \ge \frac{1}{m F_Q (T) },
\end{equation}
where $F_Q(T)$ is the QFI (we set $k_{\rm B}=1$). In the scenario of a parameter dependent Hamiltonian $\hat{H}(\lambda)$ the quantum states are given by a density matrix $\hat{\rho}(T, \lambda)$, and the QFI, with respect to the temperature $T$ for a given value of the control parameter $\lambda$, is defined as the maximization of the classical Fisher information (CFI) $F_c(T,\hat{A})$ over all possible measurements~\cite{Braunstein1994}
\begin{equation}
    F_Q(T)=\underset{\hat{A}}{\rm max}\,F_c(T,\hat{A}),
\end{equation}
with $\hat{A}$ representing any possible observable of the state, and
\begin{equation}\label{eq:FC}
    F_c(T,\hat{A}) = \sum_\alpha \frac{1}{p(\alpha|T)} \left( \frac{\partial p(\alpha|T) }{\partial (T)} \right)^2,
\end{equation}
where $p(\alpha|T)=\langle \alpha|\hat{\rho}(T,\lambda)|\alpha\rangle$ (and $\hat{A}|\alpha\rangle = \alpha|\alpha\rangle$) is a conditional probability distribution of the measurement outcomes for $\hat{A}$, given a fixed value of temperature $T$~\cite{Paris2015, Mehboudi2019_1}.

It has been proved that the optimal measurement (i.e., the measurement which gives $F_c(T,\hat{A}) = F_Q(T)$) is the energy of the system~\cite{Paris2015, Salvatori2014}, i.e., $\hat{A} = \hat{H}(\lambda)$ with eigenvalues $\hat{H}(\lambda)|\psi_n\rangle = E_n(\lambda)|\psi_n\rangle$. Therefore it is convenient to define the energy gap between $n-$th excited and ground state as follows: $\Delta_n(\lambda) = E_n(\lambda) - E_0(\lambda)$. Considering a state well described within the canonical Gibbs ensemble
\begin{equation}\label{rho}
    \hat{\rho}(T,\lambda)=\sum_{n=0}^{\mathcal{N}} \frac{e^{-\Delta_{n}(\lambda)/T}}{Z} |\psi_n\rangle \langle \psi_{n}|,
\end{equation}
where $\mathcal{N}$ is the total number of quantum states, $Z$ is the partition function, and applying Eq.~\eqref{eq:FC}, one obtains
\begin{equation}\label{eq:fisher}
    F_Q(T,\lambda) = \frac{\Delta^2 \hat{H}(T, \lambda)}{T^4},
\end{equation}
where $\Delta^2$ denotes the standard definition for the variance of an operator, $\Delta^2 \hat{A} \equiv \langle \hat{A}^2 \rangle - \langle \hat{A} \rangle^2$. The relation~\eqref{eq:fisher} can also be obtained by using the fidelity susceptibility approach~\cite{Mehboudi2019_1} as conveniently used in the estimation of the parameter of the Hamiltonian~\cite{Zanardi2007_2}. The value of the QFI depends on the structure of energy levels and the temperature of a given quantum system. Note that in the remainder of the work, unless required, we drop the explicit dependence of $F_Q$ and $F_c$ with the observable $\hat{A}$ and the parameters $\lambda$ and $T$.

\subsection{Finite-size scaling at continuous quantum phase transitions}
We consider a $d$-dimensional quantum many-body system of size $L^{d}$ consisting of $N$ particles and described by the Hamiltonian $\hat{H}(\lambda)=\hat{H}_0+\lambda \hat{V}$, where $[\hat{H}_0,\hat{V}]\ne 0$ and $\lambda$ is a dimensionless control parameter. We assume the QPT takes place in the thermodynamic limit ($N\to \infty$, $L^{d}\to\infty$, $N/L^{d}={\rm const.}$) at zero temperature and is driven by $\lambda$ with the critical point located at $\lambda=\lambda_{c}$. We therefore define $\epsilon \equiv \lambda-\lambda_c$ as the distance from the critical point. The QPT is characterized by a diverging power law behavior of a physical quantity $A \sim \epsilon^{a}$ given by a critical exponent $a$ quantifying how rapidly $A$ changes at $\lambda_c$, for example, if $A$ is the energy gap between the ground and first excited energy states $\Delta_g \equiv \Delta_1(\epsilon)$ then $\Delta_g\sim \epsilon^{-z\nu}$~\cite{Continentino2001}, where $z$ is the dynamic critical exponent and $\nu$ is the critical exponent associated with the divergence of the correlation length.

The singular behavior of $A$ is observed in the thermodynamic limit. If the size of the system ($L$ and $N$) is finite then the change of $A$ is an analytic function of $\epsilon$. Moreover, around the transition point physical quantities are subject to finite-size scaling which depends on the general properties of the transition~\cite{Cardy1988, Sondhi1997, Campostrini2014}. This was understood by the finite-size scaling approach, which was also considered through the renormalization group~\cite{Cardy1988, Pelissetto2002, Fisher1972, Botet1983, Campostrini2014, Rossini2018}. In the finite-size scaling framework, an existence of a regular function $f_a$ is postulated such that
\begin{equation}\label{eq:scalingA}
    A\sim L^{-a/\nu} f_a(\epsilon L^{1/\nu})
\end{equation}
with the constraint $f_a(0)\ne 0$~\cite{Cardy1988, Campostrini2014, Rossini2018, Lacki2017}. Note, the finite-size behavior of $A$ is characterized by the scaling function and scaling laws, with critical exponents determined by the universality class of the QPT. Therefore, the behavior of $A$ in the critical region depends on global properties, such as $L, \, N,\, d$ and the symmetry or nature of the interactions~\cite{Continentino2001, Rossini2018}. The finite-size scaling framework can also be considered in terms of $N$ instead of $L$ as $L\propto N^{1/d}$. For example, if $A=\Delta_g$ and $a=z \nu$, the finite-size scaling for the energy gap~\cite{Campostrini2014} then reads
\begin{equation}\label{eq:energygapscaling}
    \Delta_g \sim N^{-z/d} f_\Delta(\epsilon N^{1/(\nu d)}).
\end{equation}

A non-zero value of temperature $T$ introduces an additional relevant parameter, which, within the finite-size scaling framework, is taken into account by adding a further dependence of the scaling functions on $T/\Delta_g$~\cite{Continentino2001, Rossini2018}.

\subsubsection{Finite-size scaling of the QFI}
We now apply the finite-size scaling approach to the thermometry problem at the QPT in order to relate the behavior of the QFI with critical exponents and relevant scaling functions. According to the discussion above we can expect that the QFI is also subject to the finite-size scaling relation~\eqref{eq:scalingA}. However, the value of the scaling exponent $a$ is not given a-priori. To analyze the scaling of the QFI and obtain $a$, we first consider the dimensionless quantity $F_{Q} T^2=\Delta^2 \hat{H}/T^2$, which can be expressed as a function of independent, dimensionless combinations of physical parameters of the system, specifically, the product $\epsilon N^{1/(\nu d)}$ as for~\eqref{eq:energygapscaling} and the ratios $T/\Delta_n$. When the temperature is low, i.e, $T\lesssim \Delta_g$, the main contribution to the quantity $F_{Q} T^2$ comes from the terms containing $T/\Delta_g$ while the contribution of $T/\Delta_{n>1}$ is assumed to be negligible~\footnote{In fact, according to~\cite{Campostrini2014} not only the lowest energy gap $\Delta_g$ but also other lower lying energy gaps $\Delta_{n\ne 1}$ follows the scaling~\eqref{eq:energygapscaling}.}. This approximation leads to the following scaling form of the QFI near the critical point
\begin{equation}\label{eq:FQscalingA}
    F_Q = \Delta_g^{-2}\tilde{g} \left(\epsilon N^{1/(\nu d)}, \frac{T}{\Delta_g} \right),
\end{equation}
where the low temperature approximate of the scaling function $\tilde{g}$ can be expressed as $\tilde{g}\left(\epsilon N^{1/(\nu d)}, \frac{T}{\Delta_g} \right)\equiv \left(\Delta_g/T\right)^2 \left( \Delta^2 \hat{H}/T^2 \right)$ (see~\autoref{app:scaling} for details). Note, Eq.~\eqref{eq:FQscalingA} shows the direct relation between the QFI and the inverse of the energy gap squared. As a consequence, the QFI grows at the critical point at a rate determined by the speed at which the energy gap closes. Under this approximation, $\tilde{g}$ can be calculated analytically as a function of the energy gap $\Delta_g$ (see~\autoref{app:scaling}), and it is
\begin{equation}\label{eq:FQscalingF}
    \tilde{g}\left(\frac{ T}{\Delta_g}\right) = \left(\frac{\Delta_g}{T}\right)^4 \frac{1}{4 \cosh^2{\left[\frac{\Delta_g}{2 T}\right]}}.
\end{equation}
The scaling function can be seen as a single-parameter function of $T/\Delta_g$, which tends to zero when temperature is much larger and smaller than the energy gap. The condition in which $T \gg \Delta_g$ resembles a classical situation in which the temperature is much larger than many internal energy levels in addition to the lower-lying ones, thus frustrating all the information enhancement provided by the closing of the energy gap, making this limit uninteresting for our purposes. An analysis shows that the maximal value of the QFI from~\eqref{eq:FQscalingF} near the critical point is  approximately $F_Q^{\rm max} \approx 4.53\Delta_{g}^{-2}$ for $T/ \Delta_{g}\approx 0.24$. These values estimate well the position and maximum of the QFI for the strongly correlated system we are considering in the next sections. Taking the scaling of the energy gap~\eqref{eq:energygapscaling} in~\eqref{eq:FQscalingA}, we obtain that the QFI around criticality scales with the total number of particles as $N^{2z/d}$,
\begin{equation}\label{eq:FQscaling}
    F_Q \sim N^{2z/d}  \frac{\tilde{g} \left(\epsilon N^{1/(\nu d)}, \frac{T}{\Delta_g} \right)}{f^2(\epsilon N^{1/(\nu d)})},
\end{equation}
when considered for fixed values of $\epsilon N^{1/(\nu d)}$ and $T/\Delta_g$.

Until now, we have considered a general scenario for non-degenerate systems. However, degeneracy might have important consequences for the scaling of the QFI. Here we just note some simple observations. When the degeneracy is independent of the number of particles $N$ it appears as a constant multiplicative factor to the scaling function and can be neglected in the analysis of the scaling exponents. However, a more complicated scenario appears if the degeneracy of the energy levels  is a function of $N$. In this case the scaling exponents can differ from those considered in the non-degenerate scenario, see~\autoref{app:scaling} for more details. Indeed, the optimal structure of energy levels, which leads to the highest value of the QFI, involves a single ground state and all degenerated excited states with a well-defined energy gap~\cite{Correa2015, Paris2015}. The engineering of the optimal energy levels and the resulting sensitivity are summarized in~\autoref{app:optimalQFI}.

We now discuss the SNR, defined as $T/\sqrt{\delta^2 T}$, also called the relative estimation precision. This quantity is dimensionless itself and it can be related to the QFI using~\eqref{eq:CRLB}. This gives $T \sqrt{\delta^2 T}=\sqrt{F_Q}T$. From the scaling behavior of the QFI in~\eqref{eq:FQscalingA}, we can see that the SNR can be expressed as
\begin{equation}\label{eq:scaling-signal-to-noise}
    \frac{T}{\sqrt{\delta^2 T}} = \left( \frac{T}{\Delta_g}\right)\tilde{g}^{1/2}\left( \epsilon N^{1/(\nu d)}, \frac{T}{\Delta_g} \right).
\end{equation}
This implies that, as before, the SNR near the critical point is given by a scaling function of two parameters $\epsilon N^{1/(\nu d)}$ and $T/\Delta_g$. Unlike the QFI, the SNR does not exhibit the scaling with the total number of particles for fixed values of $\epsilon N^{1/(\nu d)}$ and $T/\Delta_g$.

\begin{figure*}[]
\centering
    \begin{picture}(260,240)
    \put(-120,0)
{\includegraphics[width=1\linewidth]{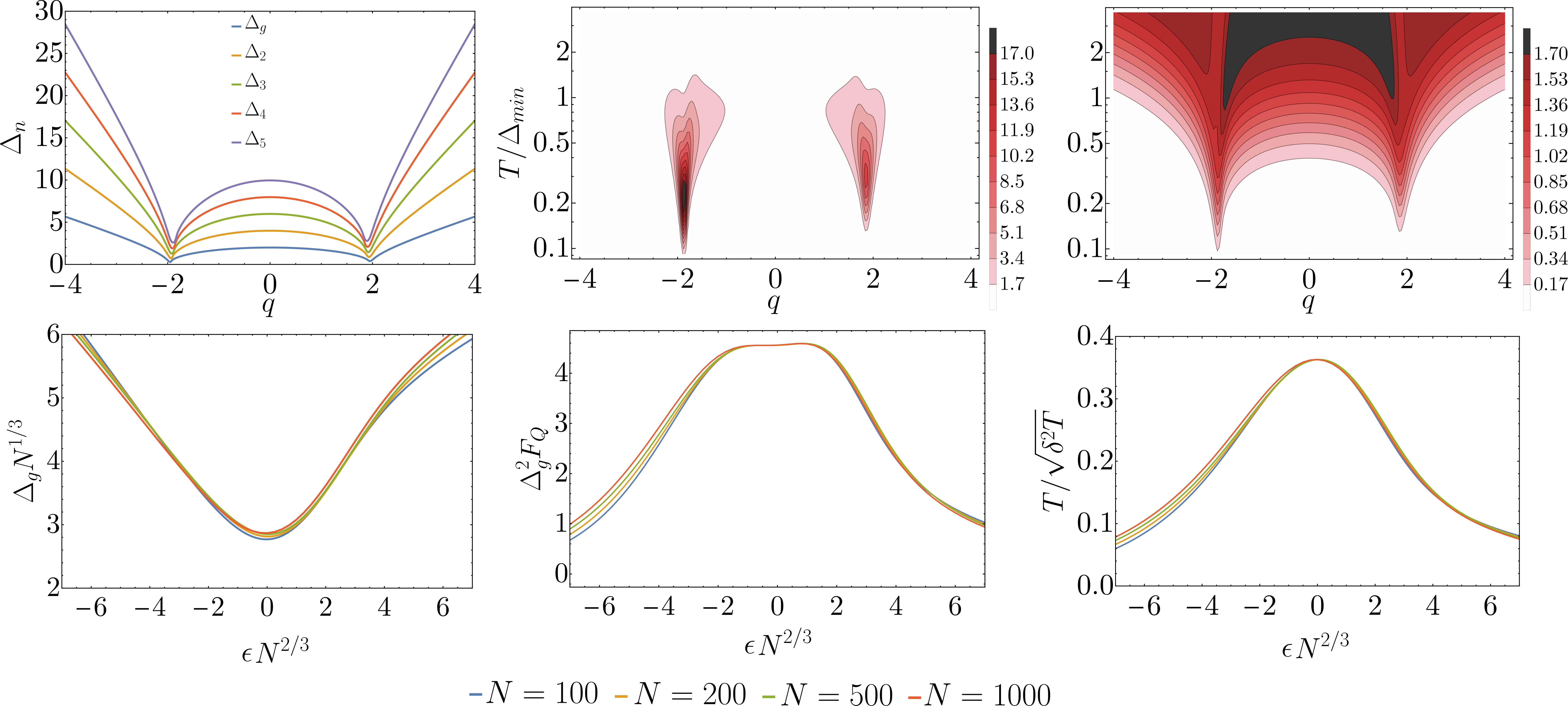}}
    \put(0,200){\small $(a)$}
    \put(165,200){\small $(b)$}
    \put(330,145){\small $(c)$}
    \put(12,102){\small $(d)$}
    \put(165,102){\small $(e)$}
    \put(330,102){\small $(f)$}
\end{picture}
\caption{$(a)$ Energy levels of $\hat{H}_1/c$ as a function of quadratic Zeeman energy $q=q_z/c$ for $N=1000$. Contour plots of the quantum Fisher information $F_Q$ in units of $c^{-2}$ $(b)$ and the signal-to-noise ratio $T/\sqrt{\delta^2 T}$ $(c)$ versus $q$ and $T/\Delta_{\rm min}$ for $N=200$. Numerical demonstration of finite-size scaling of $\Delta_g$ $(d)$, $F_Q$ $(e)$ and $T/\sqrt{\delta^2 T}$ $(f)$ versus the product $N^{2/3}\epsilon$ around the left critical point when $T/\Delta_{\rm min} = 0.25$ $(e)$, and $T/\Delta_{\rm min} = 0.17$ $(f)$, for various number of atoms $N$ as indicated in figure legend. Here, $q_c$ is defined as a position of the minimum of $\Delta_q$ and depends on $N$. In~\autoref{app:noscaling}, we show the numerical results before applying the scaling.}
\label{fig:fig1}
\end{figure*}
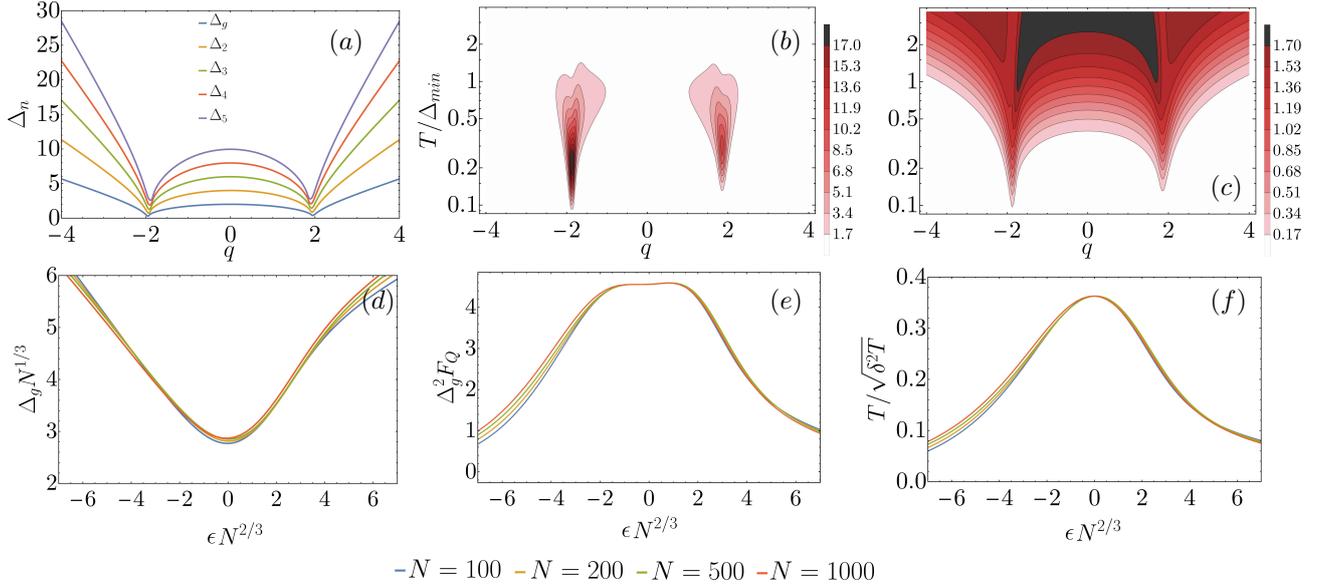

In this section, we have applied the finite-size scaling approach for critical systems to the thermometry problem. We derived an expression for the scaling function and scaling exponents of the sensitivity bound in temperature estimation  quantified by the QFI. In the following two sections, we investigate the sensitivity of critical quantum thermometry and the scaling properties around the critical region of two different systems. First, we consider the spin-1 BEC belonging to the all-connected spins universality class such as the Lipkin-Meshkov-Glick model~\cite{Lipkin1965}. Next, we do the same for the spin-1/2 XX model with nearest neighbors interaction~\cite{Rams2018}. These systems are different in the structure of their lowest energy levels. However, both of them exhibit phase transition as a function of a Hamiltonian parameter and display asymptotically closing energy gaps at the critical points.

\section{Spin-1 Bose-Einstein Condensate}\label{sec:spin1}

In this section we consider the finite size spin-1 BEC system in the $F=1$ hyperfine ground state manifold in the presence of an external homogeneous magnetic field. Under the single mode approximation (SMA) ~\cite{Kawaguchi2012, Stamper-Kurn2013, Mirkhalaf2021}, the Hamiltonian for $N$ atoms reads
\begin{equation}\label{eq:HS1}
    \hat{H}_1 = -\frac{c}{2N}\hat{J}^2 - q_{\rm z} \hat{N}_0
\end{equation}
after dropping constant terms. Here, $\hat{J}^2 = \hat{J}_x^2 + \hat{J}_y^2 + \hat{J}_z^2$ where $\hat{J}_{x,y,z}$ are the spin-1 operators, $\hat{N}_{m_F}$ is the number of atoms in the ${m_F}=0, \pm 1$ Zeeman state, $q_{\rm z}$ is the quadratic Zeeman energy shift which acts as the control parameter $\lambda$, and ${c}=N|c_2|\int d{\boldsymbol{r}}|\phi(\boldsymbol{r})|^4$, where $\phi(\boldsymbol{r})$ is the spatial atomic wave function being a solution of the Gross-Pitaevskii equation~\cite{Kawaguchi2012, Stamper-Kurn2013}~\footnote{The explicit form of $c_2$ is given by $c_2=4\pi\hbar^2(a_0 -a_2)/3m$, where $m$ is the mass of each particle and $a_0(a_2)$ is the s-wave scattering length for spin-1 atoms colliding in symmetric channels of total spin $J=0 \, (J=2)$. Note, $c$ is proportional to the density $c\propto \rho=N/V$ for homogeneous systems.}. In the following we will work with the re-scaled Hamiltonian $\hat{H}_1/c$, and we will refer to the re-scaled quadratic Zeeman energy as $q = q_{\rm z}/c$.

In general, the system can be viewed as $N$ interacting spin-1 particles exposed to an external field where the interaction is not limited to the nearest neighbors. The total magnetization $\mathcal{M}$ is defined as $\mathcal{M} \equiv N_{1} - N_{-1}$. This quantity is a constant of motion due to collisional symmetry~\cite{Kawaguchi2012}. Throughout the remainder of this work, we will only consider the case in which $N_1 = N_{-1}$, i.e., $\mathcal{M}=0$, and the variance $\Delta^2 \mathcal{M} = 0$.

In the case of $\mathcal{M}=0$, the phase diagram of ground states contains two critical points that separate the polar (P) phase (characterized by $N_{\pm 1} = 0$) from the broken axisymmetry (BA) phase ($N_0, N_{\pm 1} > 0$), and the BA phase from the antiferromagnetic (AFM) phase ($ N_{\pm 1} = N/2$)~\cite{Orenes2019}. In the thermodynamic limit, the two critical points for $\hat{H}_1/c$ are $q_c=2$ and $q_c=-2$, respectively. These critical points were shown to provide an enhanced sensitivity in the estimation of the control parameter $q$~\cite{Mirkhalaf2021}.

In the following, we will explore the sensitivity bounds for quantum thermometry with the spin-1 system by numerically evaluating the QFI~\eqref{eq:fisher} and SNR given by $T/\sqrt{\delta^2 T}=\sqrt{F_Q}T$ using the method described in~\cite{Mirkhalaf2020}. The results will be then interpreted within the finite-size scaling approach described in~\autoref{sec:theoryofquantumthermometry}, confirming the validity of this description and our derivation of the scaling behavior of the QFI and SNR.

\autoref{fig:fig1}$(a)$ shows the eigenvalue spectrum $\Delta_n(q)$ (for the five lowest energy levels) of $\hat{H}_1/c$ for $N=10^3$. This exhibits minima in the gap $\Delta_n$ near $q = \pm 2$.~\autoref{fig:fig1} $(b)$ and $(c)$ show by color the values of the QFI and SNR, respectively, both in the $(q-T/\Delta_{\rm min})$ parameter space. The QFI is largest around the critical points, and its value peaks at $T/\Delta_{\rm min} \approx 0.26$. However, while the SNR also increases around the critical points, it shows a different behavior, as discussed in~\autoref{sec:theoryofquantumthermometry}.

To confirm the validity of the finite-size scaling approach, we have to consider that the spin-1 BEC within the SMA is an example of the ``infinite'' coordinate system, where the concept of correlation length and dimension is lost. The scaling approach can be applied in this case with the exponents modified by the critical dimension~\cite{Botet1983, Continentino2001}. This was discussed for the spin-1 system in~\cite{Xue2018, Mirkhalaf2021}, where it was shown that $\Delta_g \sim N^{-1/3} f(\epsilon N^{2/3})$. Having this into account,~\autoref{fig:fig1}$(d)$ shows the result of applying the finite-size scaling to the lowest energy gap, as in~\autoref{fig:fig1}$(a)$ for different numbers of atoms. The scaling is confirmed with scaling exponents $z \nu = 1/2$, as in the Lipkin-Meshkov-Glick model and its corresponding universality class~\cite{Salvatori2014, Dusuel2004}.~\autoref{fig:fig1}~$(e)$ and $(f)$ show the result of applying the scaling to the numerical results for $F_Q$ and $T/\sqrt{\delta^2 T}$, respectively, as a function of the product $\epsilon N^{2/3}$ for fixed  $T/\Delta_g = 0.17$. The results evidence the scaling $F_Q \sim \Delta_g^{-2} \tilde{g}(\epsilon N^{2/3})$ and $T/\sqrt{\delta^2 T} \sim (T/\Delta_g) \tilde{g}^{1/2}(\epsilon N^{2/3})$ as expected from the finite-size scaling approach. Directly, this demonstrates that $F_Q \sim N^{2/3}$ when $T/\Delta_g$ is fixed, and the accuracy of the general finite-size scaling arguments used in~\autoref{sec:theoryofquantumthermometry} for the spin-1 BEC system.

It is worth mentioning that in the antiferromagnetic condensate~\cite{Evrard2020} the energy gap follows the scaling law~\eqref{eq:energygapscaling} for the zero magnetization case when the system exhibits a first order phase transition~\cite{Mirkhalaf2020}. The relevant scaling exponent suggests $F_Q\sim N^2$ for fixed $T/\Delta_g$ around criticality.

\section{Spin-1/2 XXZ System}\label{sec:xxz}

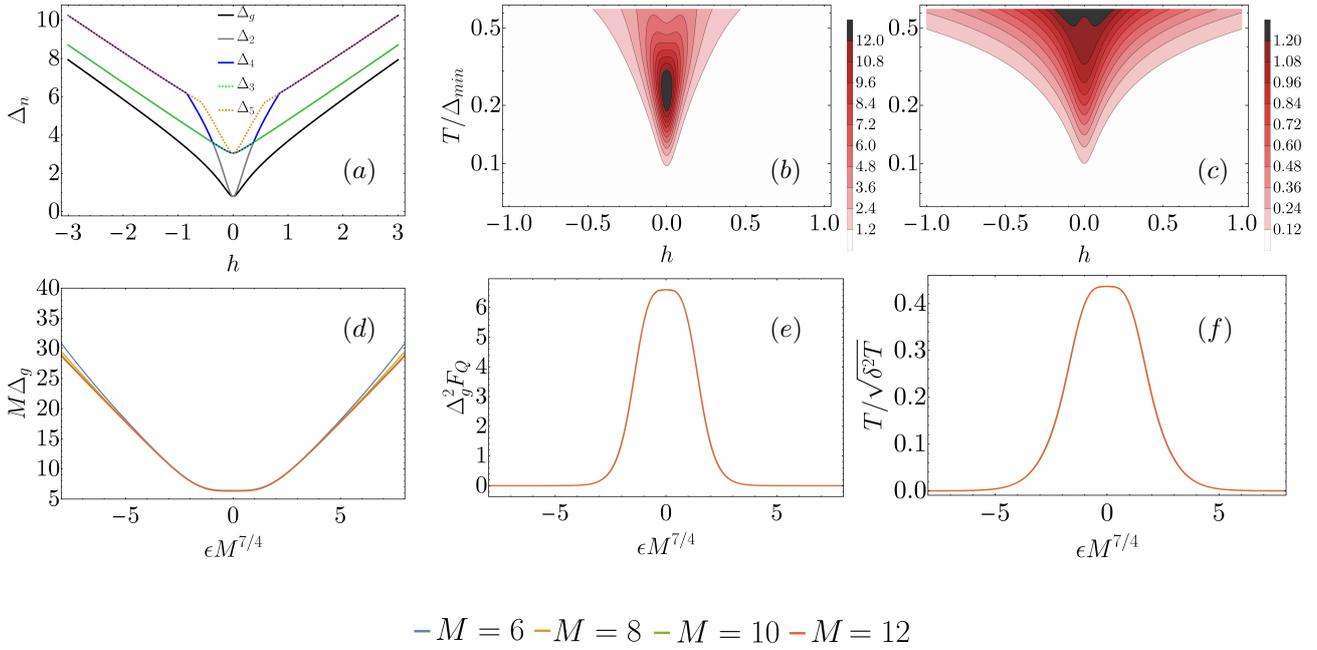
\begin{figure*}[]
\centering
    \begin{picture}(260,240)
    \put(-120,0)
{\includegraphics[width=1\linewidth]{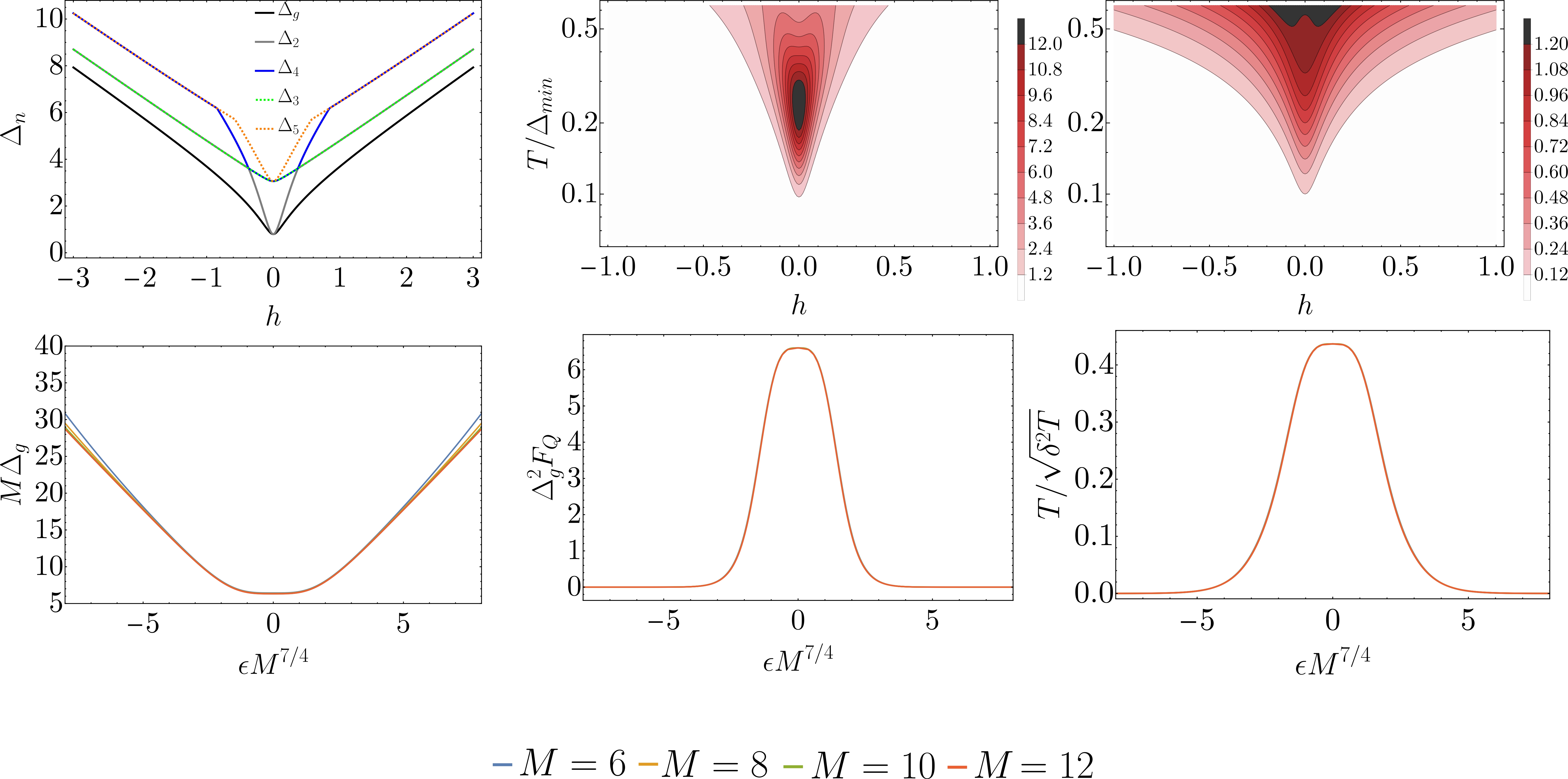}}
    \put(5,175){\small $(a)$}
    \put(165,175){\small $(b)$}
    \put(325,175){\small $(c)$}
    \put(05,115){\small $(d)$}
    \put(165,115){\small $(e)$}
    \put(325,115){\small $(f)$}
\end{picture}
\caption{$(a)$ Lowest energy level structure of $\hat{H}_{1/2}/J$ versus transverse magnetic field $h = h_x/J$ for $M=8$. Contour plot for the $F_Q$ in units of $J^{-2}$ $(b)$ and SNR $(c)$ as a function of $h_x/J$ and $T/\Delta_g$ for $M=8$. $(d)$ Energy gap $\Delta_g$ re-scaled according to~\eqref{eq:gapscalingXXZ}. Panels $(e)$ and $(f)$ display $ \Delta_g^2 F_Q$ and SNR respectively versus $\rm{sgn}(\epsilon) M \vert \epsilon \vert^{4/7}$ when $T/\Delta_{\rm min} = 0.17$ for the total number of spins as indicated in the legend. We note that the different lines in panels $(d),(e)$ and $(f)$ overlap. Thus, we show the direct numerical results before applying the finite-size scaling in~\autoref{app:noscaling}.}
\label{fig:fig2}
\end{figure*}

In this section, we discuss the XXZ model of $M$ spins in the transverse field with periodic boundary conditions described by the Hamiltonian ~\cite{Langari2004}:
\begin{align}\label{eq:Hfors1p2}
    \hat{H}_{1/2} = &  - 4 J \sum_{j=1}^M \left( \hat{s}^x_{j} \hat{s}^x_{j+1} + \hat{s}^y_{j} \hat{s}^y_{j+1} + \zeta_z \hat{s}^z_{j} \hat{s}^z_{j+1}\right) \nonumber \\ & + 2 h_x \sum_{j=1}^M \hat{s}^x_{j},
\end{align}
where $\hat{s}^{\sigma}_j$ with $\sigma = x,y,z$, are pseudo-spin-1/2 operators for $j$th site, $J$ is an exchange coupling in the XY plane, $\zeta_z$ is the exchange coupling anisotropy in the $z$ direction, and $h_x$ is the coupling strength between the system and an external, transverse, magnetic field.

In this work, we set $\zeta_z=0$, and consider the re-scaled Hamiltonian $\hat{H}_{1/2}/J$, for which the system exhibits a second-order phase transition at the critical point $h_x/J = 0$~\cite{Rams2018, Franchini2017}. The values of the relevant critical exponents are $z=1$ and $\nu=2/(4-\arccos(\zeta_z/J)/\pi $~\cite{Affleck1999}. In general, the model with $h_x\neq 0$ is not integrable due to the transverse field considered. However, in the special case of $h_x=0$, the Hamiltonian reduces to the XX model and can be solved analytically~\cite{Mikeska2004, Franchini2017, Mehboudi2015}. In what follows, we will explore the sensitivity bounds for quantum thermometry in this system evaluating the QFI~\eqref{eq:fisher} and SNR near the phase transition by performing an exact numerical diagonalization of the Hamiltonian. Defining a local temperature in spin systems is not trivial, and has led to much discussion~\cite{Hartmann2004_1, Hartmann2004_2, Garcia-Saez2009, Ferraro2012, Kliesch2019, Hernandez-Santana2015, Hernandez-Santana2021}, some of which has been confirmed in different systems~\cite{Kliesch2019, Hernandez-Santana2015, Hernandez-Santana2021}. Here, in contrast, we are considering global spin temperatures that describe the entire spin system. Our results can  nonetheless be expected to apply also locally when the boundary is sufficiently large~\cite{Hernandez-Santana2015}.

The structure of energy levels is shown in~\autoref{fig:fig2}$(a)$. Unlike the spin-1 BEC model discussed in the previous section, the spin-1/2 system features some weak degeneracy of the low energy levels around $h_x=0$. We define the distance from the critical point as $\epsilon = h_x/J$. According to~\eqref{eq:energygapscaling}, the energy gap between the two lowest energy states is expected to scale as 
\begin{equation}\label{eq:gapscalingXXZ}
    \Delta_g \sim M^{-1}f(\epsilon M^{7/4}),
\end{equation}
with the particular values of critical exponents $z \nu=4/7$. This is shown by the numerical results presented in~\autoref{fig:fig2}$(d)$, where we plot the re-scaled $\Delta_g$ according to~\eqref{eq:gapscalingXXZ} as a function of ${\rm sgn}(\epsilon) \vert\epsilon \vert M^{7/4}$.

In~\autoref{fig:fig2}$(b)$ and $(c)$, we show by color $F_Q$ from~\eqref{eq:fisher} and $T/\sqrt{\delta^2 T}$ for $\hat{H}_{1/2}/J$, quantifying the sensitivity and the SNR in the estimation of temperature in the $(h_x/J-T/\Delta_{\rm min})$ parameter space. The increase of the sensitivity at the critical region is again clearly visible when $T \lesssim \Delta_g$ as expected by the finite-size scaling approach  discussed in~\autoref{sec:theoryofquantumthermometry}.

Finally, we examine the finite-size scaling of both, the QFI and SNR versus ${\rm sgn}(\epsilon) M \vert\epsilon\vert^{4/7}$ when $T/\Delta_g$ is fixed. As we observe in~\autoref{fig:fig2}$(e)$ and~\autoref{fig:fig2}$(f)$ respectively, initially different curves, that correspond to different total number of spins, collapse into each other after the scaling is applied, when $T/\Delta_g=0.17$. Our results clearly demonstrate that the sensitivity is subject to the finite-size scaling, and $F_Q \sim \Delta_g^{-2} \tilde{g}(\epsilon M^{7/4})$ and $T/\sqrt{\delta^2 T} \sim (T/\Delta_g) \tilde{g}^{1/2}(\epsilon M^{7/4})$ applies for the XXZ model even in the presence of a weak degeneracy, and that the scaling of the QFI with the number of spins is $F_Q\sim M^2$ for fixed $T/\Delta_g$ at the critical point.

\section{Feasibility of Critical Quantum Thermometry in Spin Systems}\label{sec:experiment}

In the previous sections, we discussed the critical quantum thermometry for two examples of physical systems having different critical and scaling exponents. We also demonstrated by means of exact numerical calculations, that the finite-size scaling approach well captures universal scaling properties of the sensitivity. In addition, both systems can be realized in ultra-cold atomic experiments~\cite{Jepsen2020, Zou2018}. In this section, we discuss the feasibility of criticality-aided temperature sensing in such systems in reference to current experimental capabilities. We assess this considering different observables of the system, and quantifying the sensitivity that a possible experiment could reach using~\eqref{eq:FC} and the error propagation formula (EPF). The EPF is used in the estimation of a parameter, in this case $T$, evaluated from accessible experimental measurements of an observable $\hat{A}$, such that:
\begin{equation}\label{eq:EPF}
    \varDelta^2 T = \frac{\Delta^2 \hat{A}}{|\partial_T\langle\hat{A}\rangle|^2}.
\end{equation}

Let us start the analysis with the spin-1 system. We consider two possible measurements that can be experimentally realized: \emph{i}) non-destructive Faraday rotation (FR) of the collective spin degree of freedom $\hat{A} = \hat{J}_{\perp}^2\equiv \hat{J}^2 - \hat{J}^2_z$ \footnote{Note, because $\hat{J}^2_\perp  = \hat{J}^2 - \hat{J}_z^2$ the equivalence of ${\langle\hat{J}^2_\perp\rangle}$ with ${\langle\hat{J}^2\rangle}$ arises due to the fact that we are considering $\mathcal{M} = \langle\hat{J}_z\rangle = 0$, and the assumption of fixed (non-fluctuating) magnetization, i.e., $\Delta^2 \mathcal{M} = 0$.}~\cite{Palacios2018, Mirkhalaf2021, Gomez2020} and \emph{ii}) destructive absorption imaging of the $m_F = 0$ state population $\hat{A} = \hat{N}_0$. We note that the combination of such measurements is possible within the same experimental run since FR measurements are non-destructive and can be performed before a final destructive measurement of~$\hat{N}_0$.

\begin{figure}[]
\centering
    \begin{picture}(240,160)
    \put(0,0)
{\includegraphics[width=\linewidth]{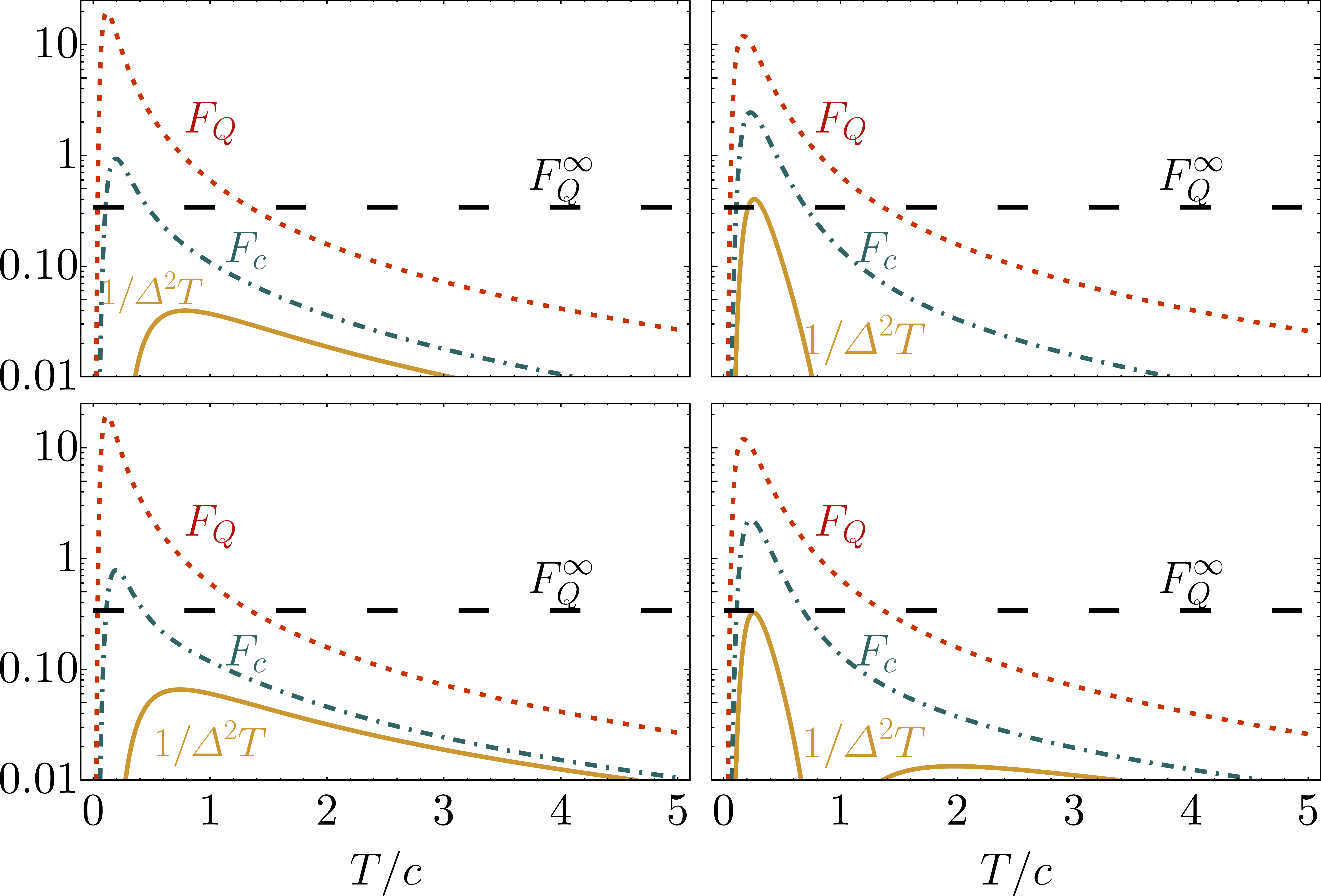}}
    \put(105,145){\small $(a)$}
    \put(217,145){\small $(b)$}
    \put(105,75){\small $(c)$}
    \put(217,75){\small $(d)$}
\end{picture}
\caption{Numerical results showing the temperature dependence of $c^2 F_Q$ (red dashed line), $c^2 F_c$ (green dot-dashed) and $c^2/\varDelta^2 T$ (orange solid) for
$\hat{J}_{\perp}^2$ $(a)$ and $(b)$, and  $\hat{N}_0$ $(c)$ and $(d)$ with $N=200$ at the critical point $q_z/c = -1.869$ $(a),(c)$, and a near-critical point $q_z/c = -1.8$ $(b),(d)$. The black-dashed lines indicate $\underset{T}{\rm max}\,F^{\rm *}_Q (q_z,T)$ to give a reference level.
}\label{fig:partial}
\end{figure}
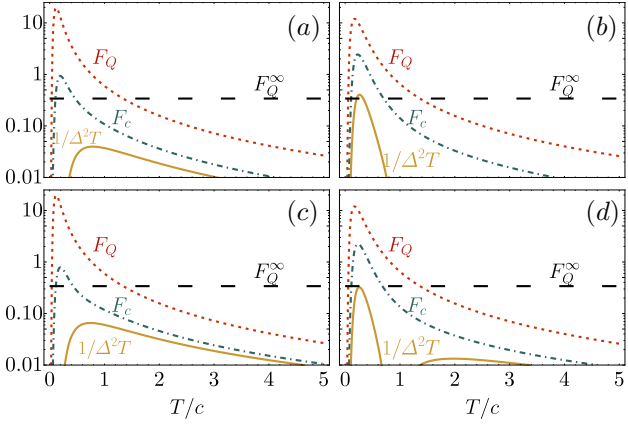

\autoref{fig:partial} shows the sensitivity in temperature estimation quantified by the $F_c$ and $1/\varDelta^2 T$ as in~\eqref{eq:FC} and~\eqref{eq:EPF} for both $\hat{A}=\{\hat{J}_{\perp}^2, \hat{N}_0\}$ in units of the energy scale $c$, versus $T/c$. As a reference, we also plot $F_Q$. Left panels $(a)$ and $(c)$ show the sensitivities at the critical point $q_z/c = -1.869$ using $\hat{J}_\perp^2$ $(a)$, and $\hat{N}_0$ $(c)$. Right panels $(b)$ and $(d)$ show the sensitivities at $q_z/c = -1.8$. As expected, at the critical point $F_Q$ shows the largest value. However, both $F_c$ and the sensitivity derived from the EPF are larger at the slightly displaced value of $q_z/c =-1.8$ than at the critical point itself. Therefore, although the optimal (energy) measurement would reach a maximum sensitivity at the critical point, the more experimentally accessible quantities show its maximum at a slightly displaced position. The black-dashed lines indicate a baseline sensitivity
\begin{equation}
    F^\infty_Q \equiv \underset{T}{\textrm{max}}\,F^{\rm *}_Q(T,q_z)
\end{equation}
evaluated at the corresponding $q_z$ for the spin-1 system, where $F^{\rm *}_Q(T,q_z)$ corresponds to the QFI calculated for the Hamiltonian in~\eqref{eq:HS1} when the interaction term is neglected. In the limit of large $N$ and very low temperatures, analytic expression can be obtained giving $F^{\rm *}_Q (T,q_z)= (q_z/ T)^4 (q_z)^{-2}\sinh^{-2}(q_z/T)$ for $q_z \ne 0$, as discussed in~\autoref{app:shot_noise}. This indicates that the critical behavior of the system enhances the sensitivity in the estimation of temperature. We can see that temperature estimation based in measurements of $\hat{N}_0,\, \hat{J}^2_\perp$ yield a similar sensitivity around the critical point, and that they are suboptimal. We note that the measurement of $\hat{J}_{\perp}^2$ gives the best sensitivity at $q_z = 0$ since at this particular value, the Hamiltonian reduces to $\hat{H}_1/c=-\hat{J}_\perp^2/(2N)$. In the case of the spin-1 system, the range of temperatures for which the system exhibit  criticality-enhanced sensitivity is found in the region of an experimentally accessible range of temperatures $\sim \textrm{pK} - \textrm{nK}$ for $N=10^3$ and $|c|=h\times17\textrm{Hz}$~\cite{Zou2018}.

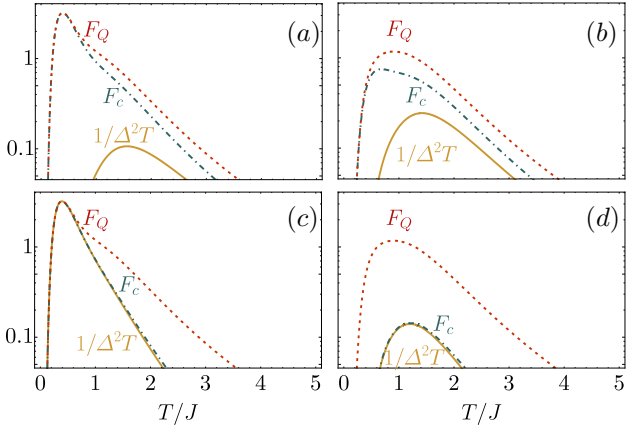
\begin{figure}[]
\centering
    \begin{picture}(240,160)
    \put(0,0)
{\includegraphics[width=\linewidth]{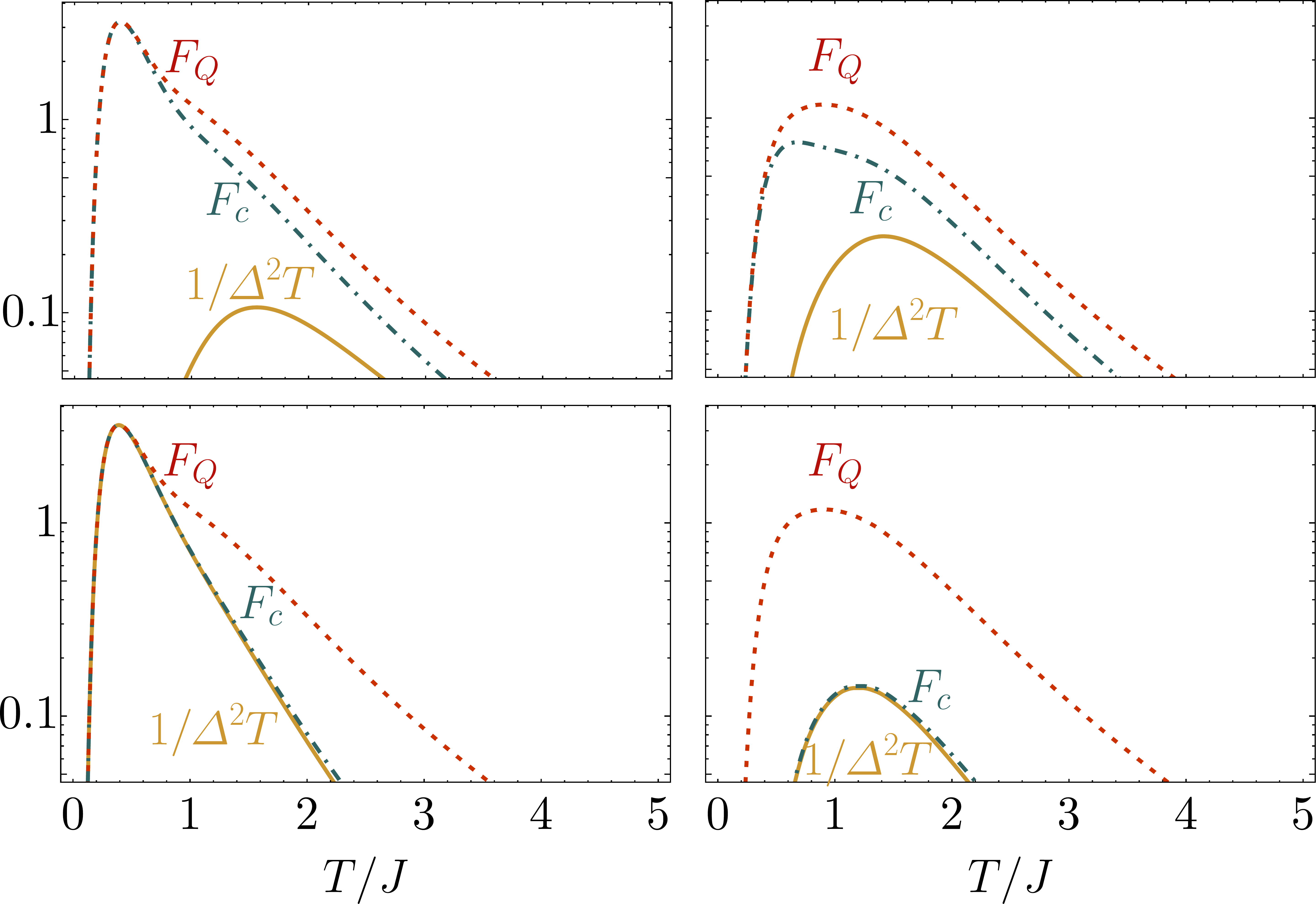}
}
    \put(105,145){\small $(a)$}
    \put(217,145){\small $(b)$}
    \put(105,75){\small $(c)$}
    \put(217,75){\small $(d)$}
\end{picture}
\caption{Numerical results showing $J^2 F_Q$ (red dashed line), $J^2 F_c$ (green dot-dashed) and $J^2/\varDelta^2 T$ (orange solid) with $M=4$ as a function of $T/J$ for $h_x/J = 0$ $(a),(c)$, and $h_x/J = 0.5$ $(b),(d)$ when $\hat{A}=\hat{S}_x^2$ $(a),(b)$, and $\hat{A}=\hat{S}_z^2$ $(c),(d)$.}
\label{fig:partialxxz_sx}
\end{figure}

The spin-1/2 XXZ model we are considering in this work has been recently realized for the first time in an ultra-cold atomic experiment~\cite{Jepsen2020} using a two-components ultra-cold bosonic gas in a 1D optical lattice. Each bosonic component takes the role of the ground(excited) state in the equivalent spin-$1/2$ model. In this case, the collective $\hat{S}_z = \sum_j \hat{s}_j^z$ and its higher moments can be measured through absorption imaging techniques, and the other collective spin components could in principle be accessed after a suitable rotation $\hat{S}_x \rightarrow \hat{S}_z$ and $\hat{S}_y \rightarrow \hat{S}_z$. More sophisticated methods, like quantum non-demolition FR measurements are also possible~\cite{Eckert2008}.

\autoref{fig:partialxxz_sx} shows the numerical results of evaluating $F_Q$, $F_c$ and $1/\varDelta^2 T$ using~\eqref{eq:fisher},~\eqref{eq:FC} and~\eqref{eq:EPF} respectively, for $\hat{H}_{1/2}/J$, and the spin observables $\hat{A}=\hat{S}_z^2$ and $\hat{A}=\hat{S}_x^2$, as a function of $T/J$.  The optimality of measuring collective quantum correlations for quantum thermometry was pointed out in~\cite{Mehboudi2015}. The results at the critical point $h_x = 0$ are shown in~\autoref{fig:partialxxz_sx} in panels $(a)$ and $(c)$. We observe that $F_c$ achieves maximum sensitivity bounded by $F_Q$ using both observables in the low temperature regime. Away from the critical point, at $h_x = 0.5$ the two signals are sub-optimal, as shown in panels $(b)$ and $(d)$. We also observe that at this value of $h_x$ and for $\hat{A}=\hat{S}_x^2$, the sensitivity obtained using the EPF equals $F_c$.

Finally, let us briefly comment on the effect of imperfect detection. Typically, there is a decrease in the sensitivity when the noise is of the order of the parameter to be measured~\cite{Len2021}. Consequently, in order to retain an acceptable sensitivity in the estimation of $T$, the detection noise should be smaller than the value of the energy gap $\Delta_g$. We briefly discuss the effect of detection noise in a particular situation in~\autoref{app:detection-noise}.

\section{Conclusions}

In this work, we have explored the critical properties of quantum thermometry in systems exhibiting continuous QPT. Using the finite-size scaling approach we linked the sensitivity in temperature estimation with universal scaling in terms of critical exponents. We showed that in the critical region and for fixed $T/\Delta_g$, the QFI scales with $N$ as $F_Q\sim N^{2z /d}$. This general relation is confirmed by exact numerical calculation of two different spin models with finite $N$: the spin-1 BEC and the Heisenberg XX spin chain which are experimentally realizable. In addition, we discussed the feasibility and optimal setup for realizing a critical quantum thermometer in such systems using practical observables. We show that measurements accessible in current experiments enable ultra-precise thermometry by exploiting critical quantum resources. The method presented here allows for both fundamental studies of thermalization processes and practical applications such as precise characterization and preparation of systems for analog quantum simulation.

\section*{Acknowledgements}

The authors kindly thank B. Damski, J. Ko\l{}ody\'nski and M. Mehboudi for fruitful discussions. EW is supported by the Polish National Science Center Grants DEC-2015/18/E/ST2/00760. AN is supported by Project no. 2017/25/Z/ST2/03039, funded by the National Science Centre, Poland, under the QuantERA programme and by MCIN/ AEI (project PID2020-115761RJ-I00). SM is supported by a grant from Basic Sciences Research, Iran, Fund (No. BSRF-phys-399-05). MWM, DBO and EA acknowledge support by the European Union from the European Metrology Programme for Innovation and Research (EMPIR) project 17FUN03-USOQS. EMPIR projects are co-funded by the European Union's Horizon 2020 research and innovation program and the EMPIR participating states. They also acknowledge H2020 QuantERA ERA-NET Cofund Q-CLOCKS (PCI2018-092973 project funded by MCIN/ AEI /10.13039/501100011033/ FEDER “A way to make Europe”), and H2020 FET Quantum Technologies Flagship project MACQSIMAL (Grant No. 820393), and the Spanish Ministry of Science projects OCARINA (PGC2018-097056-B-I00 project funded by MCIN/ AEI /10.13039/501100011033/ FEDER “A way to make Europe”) and ``Severo Ochoa'' Center of Excellence CEX2019-000910-S. Generalitat de Catalunya: CERCA, AGAUR Grant No. 2017-SGR-1354, Secretaria d'Universitats i Recerca (project QuantumCat, 
ref.~001-P-001644). Fundaci\'{o} Privada Cellex, and Fundaci\'{o} Mir-Puig.

\paragraph*{Author contributions}
AN, EA and DBO contributed equally to this work in the form of both analytic and numerical calculations. EW provided the scaling relations in~\autoref{sec:theoryofquantumthermometry}. MWM, EW and DBO conceived the idea. EW and DBO guided the research and wrote the first draft. All the authors contributed to the discussion of the results and the manuscript revision and preparation.

\appendix
\section{Optimal Configuration for the Quantum Thermometry}\label{app:optimalQFI}

The maximization of the Fisher information was already presented in the literature, e.g.,~\cite{Correa2015, Paris2015}, with the realistic proposal~\cite{Plodzien2018}. The general idea is to calculate the classical Fisher information from~\eqref{eq:FC}, which leads to the the expression
\begin{align}\label{eq:fcsum}
    F_c(T) = & \frac{1}{T^2} \frac{1}{Z^2} \sum_{\alpha=1}^m \tilde{\Delta}^2_\alpha e^{-\tilde{\Delta}_\alpha}  \nonumber \\ & + \frac{1}{T^2} \frac{1}{Z^2} \sum_{\alpha, \alpha'>\alpha} (\tilde{\Delta}_\alpha - \tilde{\Delta}_{\alpha'})^2 e^{-(\tilde{\Delta}_\alpha + \tilde{\Delta}_{\alpha'})}  \nonumber \\ \equiv & \frac{g(\tilde{\Delta}_\alpha)}{( T)^2},
\end{align}
where we introduced $g(\tilde{\Delta}_\alpha)$ to mark the whole sum in the above expression which depends on $m$ re-scaled energy gaps $\tilde{\Delta}_\alpha$.
The minimum of $F_c^{-1}$ signals the lower bound for the sensitivity in the temperature estimation. This idea is presented in~\cite{Paris2015} and later generalized for all $m$ excited states in ~\cite{Correa2015}.

When taking into account the first excited state only, $m=1$ in~\eqref{eq:fcsum}, one deals with the function $g(\tilde{\Delta}_1)$ which after maximization determines the bound for the Fisher information $F_{\rm max}(T)\approx 0.44/T^2$ (or for the temperature fluctuations $\delta^2 T_{\rm min} /T^2 \ge 2.27$).

When taking into account two excited states, $m=2$ in~\eqref{eq:fcsum}, one can easily see that the maximum for the CFI can be achieved when $\tilde{\Delta}_1=\tilde{\Delta}_2$, meaning that degeneracy of the excited states is required to obtain the maximal value. The maximal Fisher information is then $F_{\rm max}(T)\approx 0.76/ T^2$ (or for the temperature fluctuations $\delta^2 T_{\rm min} /T^2 \ge 1.31$).

Finally, one can consider all the excited states and find that the maximum of the FI is achieved when all excited states are degenerated. Then, the Fisher information $F_{\rm max}(T)\approx ({\rm log}\, m)^2/(2  T)^2$ (or for the temperature fluctuations $\delta^2 T_{\rm min} /T^2 \ge 4/({\rm log\, } m)^2$ when $m\gg 1$). More precisely, the gap that maximizes the QFI is given by:
\begin{equation}\label{eq:gapQFI}
    \frac{e^x (x-2)}{x+2}=m, \quad {\rm where}\,\, x=\Delta_{\rm max}/T
\end{equation}
and the maximal value of the QFI is
\begin{equation}\label{eq:maxQFI}
    F_{Q, \rm max}(T) =\frac{\mathcal{X}_{\rm max}}{T^2}
\end{equation}
with
\begin{equation}
    \mathcal{X}_{\rm max}=\frac{m x^2 e^{-x}}{(1 + m e^{-x})^2}.
\end{equation}

\section{Scaling Function Around Criticality}\label{app:scaling}

In this section we present a simple derivation of the relation $\Delta_g^2 F_Q$ as a function of $x=\epsilon N^{1/(\nu d)}$ and $y=T/\Delta_g$ as expressed in~\autoref{eq:FQscalingA}.

To show this, we first express the temperature in units of $\Delta_g$, namely $T=\Delta_g  (T/\Delta_g)= y \Delta_g $. Then $F_Q =\Delta_g^{-2} (\Delta^2 \hat{H})/(y^4 \Delta_g^2)$. Next, one needs to consider the quantity $(\Delta^2 \hat{H})/(\Delta_g^2)$. The variance of the energy is $\Delta^2\hat{H} = \sum_n g_n E_n^2 e^{-E_n/T}/Z - (\sum_n g_n E_n e^{-E_n/T}/Z)^2$, where $g_n$ is degeneracy of the $n$th energy level. The ground and the first excited states provide leading terms in the low temperature region ($T \lesssim \Delta_g$)~\cite{Paris2015, Zanardi2007_2}. Under this approximation the variance of the energy can be expressed as $\Delta^2\hat{H} \approx  (g_0E_0^2 + g_1 E_1^2 e^{-\Delta_g/T})/Z - (g_0E_0 + g_1 E_1  e^{-\Delta_g/T}/Z)^2$, with $Z\approx g_0+ g_1 e^{-\Delta_g/T}$, which simplifies to $\Delta^2 \hat{H} \approx g_0g_1 \Delta_g^2 /(g_0e^{\Delta_g/2T} + g_1 e^{-\Delta_g/2T})^2$ after some algebra. The ratio of energy variance and the energy gap squared then becomes $\Delta^2 \hat{H} /\Delta_g^2 \approx g_0g_1 /(g_0e^{\Delta_g/2T} + g_1 e^{-\Delta_g/2T})^2 \equiv \tilde{g} (g_0,g_1, y)$.  The dependence on $x$ arises indirectly through $y$ (since $\Delta_g = \Delta_g(x)$). The QFI can be written as
\begin{equation}
    F_Q = g_0 g_1 \Delta_g^{-2} \left(\frac{\Delta_g}{T}\right)^4 \left(g_0 e^{\Delta_g/2T} + g_1 e^{-\Delta_g/2T}\right)^{-2}
\end{equation}
Note here, that if degeneracy $g_{0/1}$ is independent of $N$ then it acts as the numerical factor and can be neglected in scaling analysis. However, if degeneracy is a function of $N$ then it affects the scaling properties of the QFI and must be taken into account.

\section{Baseline Sensitivity Based on the Uncorrelated Counterpart of the System}\label{app:shot_noise}

In this section we will consider the generic model Hamiltonian $\hat{H}^*=g\hat{H}_0$, which effectively describes the two system models considered in the main text after dropping the corresponding non-linear terms. This corresponds to the situation at which the interaction term in the Hamiltonian becomes negligible as compared to the linear term $\hat{H}^* \propto g$, and takes place in the region far from the critical point.We calculate the QFI for this generic model in its eigenbasis, assuming $\hat{H}_0 \vert n_0 \rangle = E_{n_0} \vert n_0\rangle$ with $E_{n_0} = g n_0$, $n_0=0,\cdots, \mathcal{N}$. Now, we can compute the QFI for $\hat{H}^*$ using~\eqref{eq:fisher} and obtain:
\begin{align}\label{eq:FQinfty}
    \nonumber F^{\rm *}_Q (T,g) = & T^{-4} \sum_{n_0 = 0}^\mathcal{N} \frac{E_{n_0}^2}{Z} e^{-E_{n_0}/T}\\ \nonumber & - T^{-4} \left(\sum_{n_0 = 0}^\mathcal{N} \frac{E_{n_0}}{Z} e^{-E_{n_0}/T} \right)^2 \\ \approx & \frac{1}{4 g^2} \left( \frac{g}{2 T} \right)^4 \frac{1}{ \sinh^{2}\left( g/2T\right)},
\end{align}
for $g\ne 0$, $\mathcal{N} \to \infty$ and in the limit of low temperatures, where $e^{-E_{n_0}/T}\to 0$ for large $n_0$. A careful analysis of~\eqref{eq:FQinfty} shows that its maximal possible value as a function of temperature is
\begin{equation}
    F_Q^\infty \equiv \underset{T}{\rm max}\,F_Q^{\rm *}(T,g) \approx 4.88/g^2
\end{equation}
when the temperature that maximizes the above expression $T^* \approx g/(3.83)$. This implies that the QFI tends to zero as $1/g^2$ when $g \to \infty$. If one identifies the control parameter $g$ with the distance from the critical point $\epsilon$, then one can conclude that the QFI tends to zero as $1/\epsilon^2$.

For the spin-1 system, we can  identify $g=-q_z$ and $\hat{H}_0= \hat{N}_0$. The corresponding eigenbasis is given by the Fock states $\vert n_0 \rangle = \vert k, N-2k,k \rangle$ with $k=0, \cdots, N$ and $E_{n_0}=-2 q c k$.

In particular, away from the critical point the analytic approximate expression for the QFI in the large $\mathcal{N}$ limit is
\begin{equation}
    F^{\rm *}_Q (T,q_z)= \frac{1}{q_z^2} \left( \frac{q_z}{T} \right)^4 \frac{1}{ \sinh^{2}\left( q_z/T \right)}.
\end{equation}

A similar expression can be obtained for the XXZ model. However, since in this case the critical point is $h_x = 0$, we do not use it to reference a baseline for the sensitivity in the main text.
\begin{equation}\label{eq:FQinftyXXZ}
    F^{\rm *}_Q (T,h_x)= \frac{4}{h_x^2} \left( \frac{h_x}{2 T} \right)^4 \frac{1}{ \sinh^{2}\left( h_x/2 T \right)}
\end{equation}
for the XXZ model.

\begin{figure}[ht]
\centering
\includegraphics[width=0.8\linewidth]{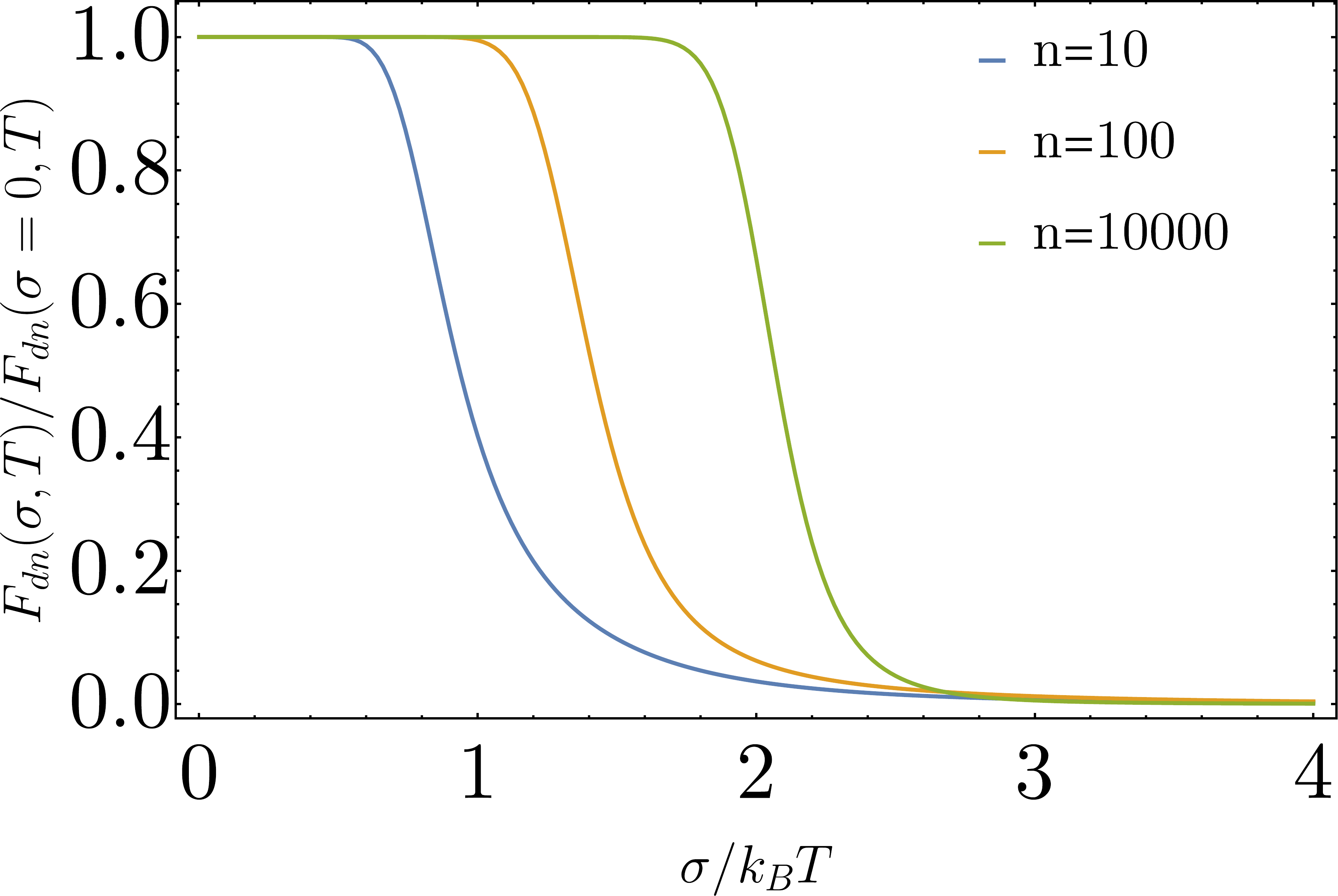}
\caption{The Fisher information with detection noise with respect to unperturbed case for the optimal case with degenerate $m$ excited states when $\Delta_1/ T = \log(m)$.}
\label{fig:figapp}
\end{figure}

\section{Effect of Detection Noise}\label{app:detection-noise}

Assuming the measurement of energy $\hat{A}=\hat{H}$, we show here the effect of detection noise. Therefore, we consider the Fisher information as in~\eqref{eq:FC} where the probability distribution $p(E_\alpha)=e^{-E_\alpha}/Z$ is replaced with
\begin{equation}
    P_{dn}(E_\alpha)=\sum_{E_{\alpha'}} \frac{e^{-(E_\alpha - E_{\alpha'})^2/2 \sigma^2}}{\mathcal{N}_{\alpha'}} P(E_{\alpha'}),
\end{equation}
with
\begin{equation}
    \mathcal{N}_{\alpha '}=\sum_{n} e^{-(E_n - E_{\alpha'})^2/2\sigma^2} .
\end{equation}

In the optimal case, considered in~\autoref{app:optimalQFI}, with all degenerated $m$ excited states one can obtain an analytical formula for the classical Fisher information 

\begin{equation}
    F_{dn} =\frac{1}{(T)^2}\frac{m(1-y)^2 \log^2m}{4 (1+y+2my )(m+2y + m y^{\log^2 m})},
\end{equation}

where we have taken the optimal energy gap $\Delta_g/T = \log(m)$ and introduced $\log y = ( T \log m/\sigma)^2$. One can easily note that in this optimal configuration the Fisher information is one parameter dependent function of dimensionless parameter $\sigma/ T$. In general, FI with detection noise can be considered as unperturbed QFI multiplied by the function of the only parameter $\sigma/ T$. A similar conclusion was made in ~\cite{Len2021}. An example of $F_{dn}$ is given in~\autoref{fig:figapp}.

\section{Numerical Results for Non-Scaled Quantities in~\autoref{fig:fig1} and~\autoref{fig:fig2}}\label{app:noscaling}

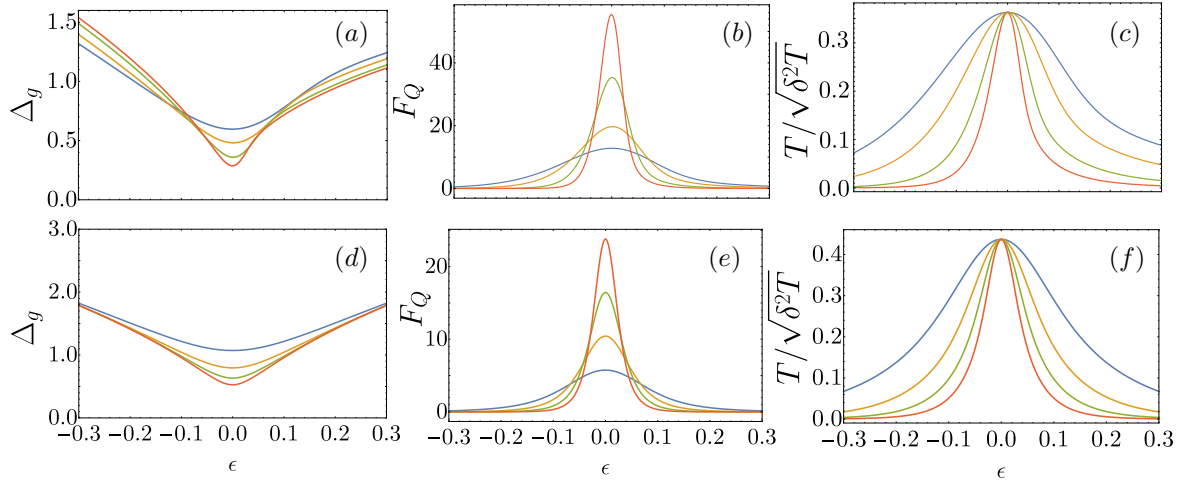
\begin{figure*}
\begin{picture}(260,200)
\put(20,0)
{
\includegraphics[width=0.9\linewidth]{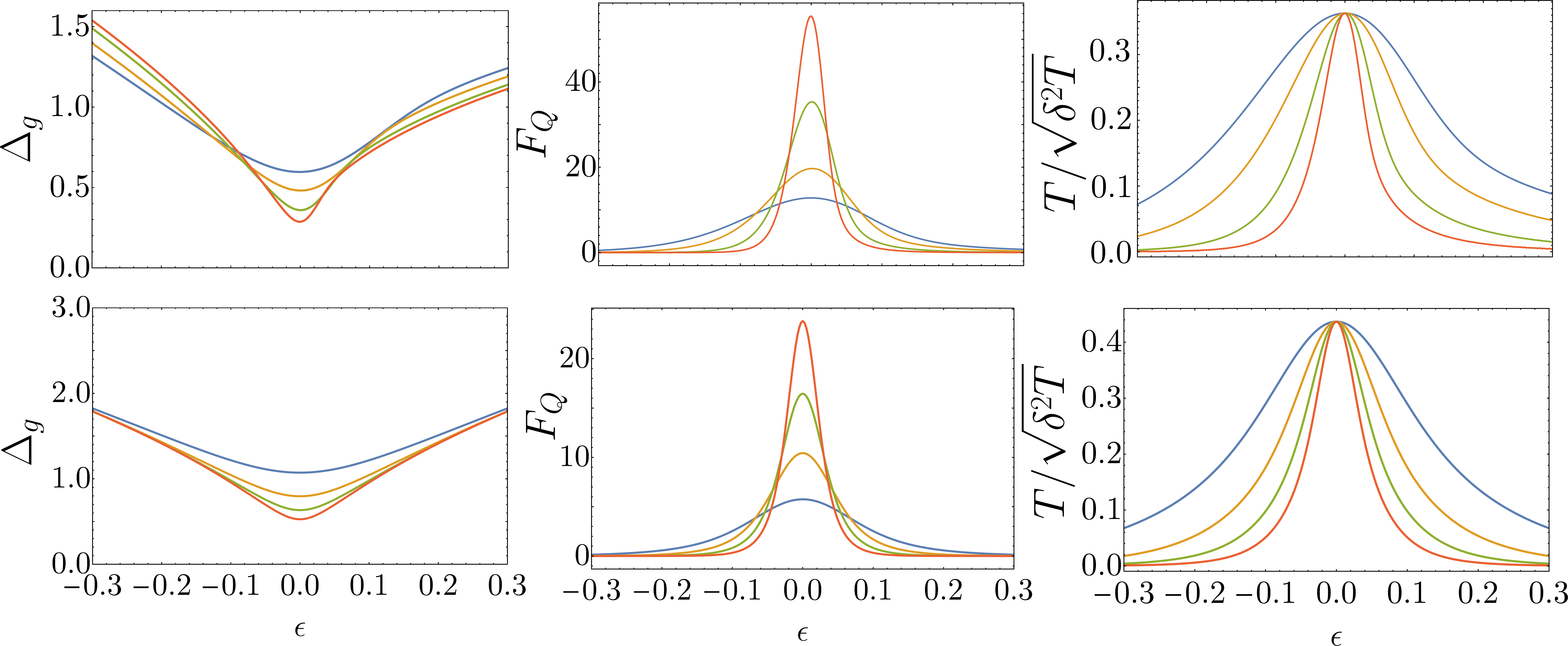}}
    \put(145,165){\small $(a)$}
    \put(290,165){\small $(b)$}
    \put(435,165){\small $(c)$}
    \put(145,80){\small $(d)$}
    \put(285,80){\small $(e)$}
    \put(435,80){\small $(f)$}
\end{picture}
\caption{Upper row shows results for the spin-1 systems. Lower row displays the corresponding quantities for the XXZ model. $(a)$ Energy gap $\Delta_g$ of the BEC spin-1 system. Panels $(b)$ and $(c)$ display $ \Delta_g^2 F_Q$ and SNR respectively versus the distance from the critical point $\epsilon$, around the left critical point when $T/\Delta_{\rm min} = 0.17$. Colors label different number of atoms $N$ as indicated in the legend of~\autoref{fig:fig1}. $(d)$ Energy gap $\Delta_g$ of the XXZ model. Panels $(e)$ and $(f)$ display $ \Delta_g^2 F_Q$ and SNR respectively versus $ \epsilon$ when $T/\Delta_{\rm min} = 0.17$. The colors indicate the total number of spins as shown in the legend of~\autoref{fig:fig2}.}
\label{fig:nonoverlapping}
\end{figure*}

\autoref{fig:nonoverlapping} shows the numerical results for the same quantities shown in panels $(d),(e)$ and $(f)$ for both~\autoref{fig:fig1} and~\autoref{fig:fig2} before applying the corresponding scaling. 

\bibliographystyle{quantum}
\bibliography{thermoquantumbiblio}

\begin{thebibliography}{10}

\bibitem{Helstrom1969}
Carl~W. Helstrom.
\newblock ``Quantum detection and estimation theory''.
\newblock \href{https://dx.doi.org/10.1007/BF01007479}{Journal of Statistical
  Physics {\bf 1}, 231--252}~(1969).

\bibitem{GobelSiegner2015}
E.~O. G\"obel and U.~Siegner.
\newblock ``Quantum metrology: {F}oundation of {U}nits and {M}easurements''.
\newblock \href{https://dx.doi.org/10.1002/9783527680887}{Wiley-VCH}. ~(2015).

\bibitem{Braunstein1994}
Samuel~L. Braunstein and Carlton~M. Caves.
\newblock ``Statistical distance and the geometry of quantum states''.
\newblock \href{https://dx.doi.org/10.1103/PhysRevLett.72.3439}{Physical Review
  Letters {\bf 72}, 3439--3443}~(1994).

\bibitem{Taddei2013}
M.~M. Taddei, B.~M. Escher, L.~Davidovich, and R.~L. de~Matos~Filho.
\newblock ``Quantum {S}peed {L}imit for {P}hysical {P}rocesses''.
\newblock \href{https://dx.doi.org/10.1103/PhysRevLett.110.050402}{Physical
  Review Letters {\bf 110}, 050402}~(2013).
\newblock  \href{http://arxiv.org/abs/1209.0362}{arXiv:1209.0362}.

\bibitem{Toth2014}
G{\'{e}}za T{\'{o}}th and Iagoba Apellaniz.
\newblock ``Quantum metrology from a quantum information science perspective''.
\newblock \href{https://dx.doi.org/10.1088/1751-8113/47/42/424006}{Journal of
  Physics A: Mathematical and Theoretical {\bf 47}, 424006}~(2014).
\newblock  \href{http://arxiv.org/abs/1405.4878}{arXiv:1405.4878}.

\bibitem{Pezze2014}
Luca Pezz{\'{e}} and Augusto Smerzi.
\newblock ``Quantum theory of phase estimation''~(2014).
\newblock  \href{http://arxiv.org/abs/1411.5164}{arXiv:1411.5164}.

\bibitem{Napolitano2011}
M.~Napolitano, M.~Koschorreck, B.~Dubost, N.~Behbood, R.~J. Sewell, and M.~W.
  Mitchell.
\newblock ``Interaction-based quantum metrology showing scaling beyond the
  {H}eisenberg limit''.
\newblock \href{https://dx.doi.org/10.1038/nature09778}{Nature {\bf 471},
  486--489}~(2011).
\newblock  \href{http://arxiv.org/abs/1012.5787}{arXiv:1012.5787}.

\bibitem{Zanardi2008}
Paolo Zanardi, Matteo G.~A. Paris, and Lorenzo Campos~Venuti.
\newblock ``Quantum criticality as a resource for quantum estimation''.
\newblock \href{https://dx.doi.org/10.1103/PhysRevA.78.042105}{Physical Review
  A {\bf 78}, 042105}~(2008).
\newblock  \href{http://arxiv.org/abs/0708.1089}{arXiv:0708.1089}.

\bibitem{Mok2021}
Wai-Keong Mok, Kishor Bharti, Leong-Chuan Kwek, and Abolfazl Bayat.
\newblock ``Optimal probes for global quantum thermometry''.
\newblock \href{https://dx.doi.org/10.1038/s42005-021-00572-w}{Communications
  Physics {\bf 4}, 62}~(2021).
\newblock  \href{http://arxiv.org/abs/2010.14200}{arXiv:2010.14200}.

\bibitem{Gietka2021}
Karol Gietka, Friederike Metz, Tim Keller, and Jing Li.
\newblock ``Adiabatic critical quantum metrology cannot reach the {H}eisenberg
  limit even when shortcuts to adiabaticity are applied''.
\newblock \href{https://dx.doi.org/10.22331/q-2021-07-01-489}{{Quantum} {\bf
  5}, 489}~(2021).
\newblock  \href{http://arxiv.org/abs/2103.12939}{arXiv:2103.12939}.

\bibitem{Chu2021}
Yaoming Chu, Shaoliang Zhang, Baiyi Yu, and Jianming Cai.
\newblock ``Dynamic {F}ramework for {C}riticality-{E}nhanced {Q}uantum
  {S}ensing''.
\newblock \href{https://dx.doi.org/10.1103/PhysRevLett.126.010502}{Physical
  Review Letters {\bf 126}, 010502}~(2021).
\newblock  \href{http://arxiv.org/abs/2008.11381}{arXiv:2008.11381}.

\bibitem{Garbe2020}
Louis Garbe, Matteo Bina, Arne Keller, Matteo G.~A. Paris, and Simone
  Felicetti.
\newblock ``{C}ritical {Q}uantum {M}etrology with a {F}inite-{C}omponent
  {Q}uantum {P}hase {T}ransition''.
\newblock \href{https://dx.doi.org/10.1103/PhysRevLett.124.120504}{Physical
  Review Letters {\bf 124}, 120504}~(2020).
\newblock  \href{http://arxiv.org/abs/1910.00604}{arXiv:1910.00604}.

\bibitem{Rams2018}
Marek~M. Rams, Piotr Sierant, Omyoti Dutta, Pawe\l{} Horodecki, and Jakub
  Zakrzewski.
\newblock ``At the {L}imits of {C}riticality-{B}ased {Q}uantum {M}etrology:
  {A}pparent {S}uper-{H}eisenberg {S}caling {R}evisited''.
\newblock \href{https://dx.doi.org/10.1103/PhysRevX.8.021022}{Physical Review X
  {\bf 8}, 021022}~(2018).
\newblock  \href{http://arxiv.org/abs/1702.05660}{arXiv:1702.05660}.

\bibitem{Mirkhalaf2020}
Safoura~S. Mirkhalaf, Emilia Witkowska, and Luca Lepori.
\newblock ``Supersensitive quantum sensor based on criticality in an
  antiferromagnetic spinor condensate''.
\newblock \href{https://dx.doi.org/10.1103/PhysRevA.101.043609}{Physical Review
  A {\bf 101}, 043609}~(2020).
\newblock  \href{http://arxiv.org/abs/1912.02418}{arXiv:1912.02418}.

\bibitem{Mirkhalaf2021}
Safoura~S. Mirkhalaf, Daniel Benedicto~Orenes, Morgan~W. Mitchell, and Emilia
  Witkowska.
\newblock ``Criticality-enhanced quantum sensing in ferromagnetic bose-einstein
  condensates: Role of readout measurement and detection noise''.
\newblock \href{https://dx.doi.org/10.1103/PhysRevA.103.023317}{Physical Review
  A {\bf 103}, 023317}~(2021).
\newblock  \href{http://arxiv.org/abs/2010.13133}{arXiv:2010.13133}.

\bibitem{Pezze2019}
Luca Pezz{\'{e}}, Andreas Trenkwalder, and Marco Fattori.
\newblock ``Adiabatic {S}ensing {E}nhanced by {Q}uantum {C}riticality''~(2019).
\newblock  \href{http://arxiv.org/abs/1906.01447}{arXiv:1906.01447}.

\bibitem{Salvatori2014}
Giulio Salvatori, Antonio Mandarino, and Matteo G.~A. Paris.
\newblock ``Quantum metrology in {L}ipkin-{M}eshkov-{G}lick critical systems''.
\newblock \href{https://dx.doi.org/10.1103/PhysRevA.90.022111}{Physical Review
  A {\bf 90}, 022111}~(2014).
\newblock  \href{http://arxiv.org/abs/1406.5766}{arXiv:1406.5766}.

\bibitem{Tsang2013}
Mankei Tsang.
\newblock ``Quantum transition-edge detectors''.
\newblock \href{https://dx.doi.org/10.1103/PhysRevA.88.021801}{Physical Review
  A {\bf 88}, 021801}~(2013).
\newblock  \href{http://arxiv.org/abs/1305.1750}{arXiv:1305.1750}.

\bibitem{Zanardi2007_1}
Paolo Zanardi, H.T. Quan, Xiaoguang Wang, and C.P. Sun.
\newblock ``Mixed-state fidelity and quantum criticality at finite
  temperature''.
\newblock \href{https://dx.doi.org/10.1103/PhysRevA.75.032109}{Physical Review
  A {\bf 75}, 032109}~(2007).
\newblock
  \href{http://arxiv.org/abs/quant-ph/0612008}{arXiv:quant-ph/0612008}.

\bibitem{You2007}
Wen-Long You, Ying-Wai Li, and Shi-Jian Gu.
\newblock ``Fidelity, dynamic structure factor, and susceptibility in critical
  phenomena''.
\newblock \href{https://dx.doi.org/10.1103/PhysRevE.76.022101}{Physical Review
  E {\bf 76}, 022101}~(2007).
\newblock
  \href{http://arxiv.org/abs/quant-ph/0701077}{arXiv:quant-ph/0701077}.

\bibitem{Hauke2016}
Philipp Hauke, Markus Heyl, Luca Tagliacozzo, and Peter Zoller.
\newblock ``Measuring multipartite entanglement through dynamic
  susceptibilities''.
\newblock \href{https://dx.doi.org/10.1038/nphys3700}{Nature Physics {\bf 12},
  778--782}~(2016).
\newblock  \href{http://arxiv.org/abs/1509.01739}{arXiv:1509.01739}.

\bibitem{Gu2010}
Shi-Jian Gu.
\newblock ``Fidelity approach to quantum phase transitions''.
\newblock \href{https://dx.doi.org/10.1142/s0217979210056335}{International
  Journal of Modern Physics B {\bf 24}, 4371–4458}~(2010).
\newblock  \href{http://arxiv.org/abs/0811.3127}{arXiv:0811.3127}.

\bibitem{Ashida2018}
Yuto Ashida, Keiji Saito, and Masahito Ueda.
\newblock ``Thermalization and {H}eating {D}ynamics in {O}pen {G}eneric
  {M}any-{B}ody {S}ystems''.
\newblock \href{https://dx.doi.org/10.1103/physrevlett.121.170402}{Physical
  Review Letters{\bf 121}}~(2018).
\newblock  \href{http://arxiv.org/abs/1807.00019}{arXiv:1807.00019}.

\bibitem{Ivanov2019}
Peter~A. Ivanov.
\newblock ``Quantum thermometry with trapped ions''.
\newblock \href{https://dx.doi.org/10.1016/j.optcom.2018.12.013}{Optics
  Communications {\bf 436}, 101--107}~(2019).
\newblock  \href{http://arxiv.org/abs/1809.01451}{arXiv:1809.01451}.

\bibitem{Vennettilli2021}
Michael Vennettilli, Soutick Saha, Ushasi Roy, and Andrew Mugler.
\newblock ``Precision of protein thermometry''.
\newblock \href{https://dx.doi.org/10.1103/PhysRevLett.127.098102}{Physical
  Review Letters {\bf 127}, 098102}~(2021).
\newblock  \href{http://arxiv.org/abs/2012.02918}{arXiv:2012.02918}.

\bibitem{Continentino2001}
M.~A. Continentino.
\newblock ``Quantum scaling in many-body systems''.
\newblock \href{https://dx.doi.org/10.1017/CBO9781316576854}{World Scientific
  Publishing, Singapore}. ~(2001).

\bibitem{Cardy1988}
J.~Cardy, editor.
\newblock ``Finite-size scaling''.
\newblock Elsevier Science Publisher, Amsterdam: North Holland. ~(1988).
\newblock
  url:~\href{https://www.elsevier.com/books/finite-size-scaling/cardy/978-0-444-87109-1}{www.elsevier.com/books/finite-size-scaling/cardy/978-0-444-87109-1}.

\bibitem{Campostrini2014}
Massimo Campostrini, Andrea Pelissetto, and Ettore Vicari.
\newblock ``Finite-size scaling at quantum transitions''.
\newblock \href{https://dx.doi.org/10.1103/physrevb.89.094516}{Physical Review
  B\,\,{\bf 89}}~(2014).
\newblock  \href{http://arxiv.org/abs/1401.0788}{arXiv:1401.0788}.

\bibitem{Zanardi2007_2}
Paolo Zanardi, Paolo Giorda, and Marco Cozzini.
\newblock ``Information-{T}heoretic {D}ifferential {G}eometry of {Q}uantum
  {P}hase {T}ransitions''.
\newblock \href{https://dx.doi.org/10.1103/PhysRevLett.99.100603}{Physical
  Review Letters {\bf 99}, 100603}~(2007).

\bibitem{Zanardi2007_3}
Paolo Zanardi, Lorenzo Campos~Venuti, and Paolo Giorda.
\newblock ``Bures metric over thermal state manifolds and quantum
  criticality''.
\newblock \href{https://dx.doi.org/10.1103/PhysRevA.76.062318}{Physical Review
  A {\bf 76}, 062318}~(2007).
\newblock  \href{http://arxiv.org/abs/0707.2772}{arXiv:0707.2772}.

\bibitem{Zou2018}
Yi-Quan Zou, Ling-Na Wu, Qi~Liu, Xin-Yu Luo, Shuai-Feng Guo, Jia-Hao Cao,
  Meng~Khoon Tey, and Li~You.
\newblock ``Beating the classical precision limit with spin-1 dicke states of
  more than 10,000 atoms''.
\newblock
  \href{https://dx.doi.org/https://doi.org/10.1073/pnas.1715105115}{Proceedings
  of the National Academy of Sciences {\bf 115}, 6381--6385}~(2018).
\newblock  \href{http://arxiv.org/abs/1802.10288}{arXiv:1802.10288}.

\bibitem{Jepsen2020}
Paul~Niklas Jepsen, Jesse Amato-Grill, Ivana Dimitrova, Wen~Wei Ho, Eugene
  Demler, and Wolfgang Ketterle.
\newblock ``Spin transport in a tunable heisenberg model realized with
  ultracold atoms''.
\newblock \href{https://dx.doi.org/10.1038/s41586-020-3033-y}{Nature {\bf 588},
  403--407}~(2020).
\newblock  \href{http://arxiv.org/abs/2005.09549}{arXiv:2005.09549}.

\bibitem{Hohmann2016}
Michael Hohmann, Farina Kindermann, Tobias Lausch, Daniel Mayer, Felix Schmidt,
  and Artur Widera.
\newblock ``Single-atom thermometer for ultracold gases''.
\newblock \href{https://dx.doi.org/10.1103/PhysRevA.93.043607}{Physical Review
  A {\bf 93}, 043607}~(2016).
\newblock  \href{http://arxiv.org/abs/1601.06067}{arXiv:1601.06067}.

\bibitem{Bouton2020}
Quentin Bouton, Jens Nettersheim, Daniel Adam, Felix Schmidt, Daniel Mayer,
  Tobias Lausch, Eberhard Tiemann, and Artur Widera.
\newblock ``Single-atom quantum probes for ultracold gases boosted by
  nonequilibrium spin dynamics''.
\newblock \href{https://dx.doi.org/10.1103/PhysRevX.10.011018}{Physical Review
  X {\bf 10}, 011018}~(2020).

\bibitem{Leanhardt2003}
A.E. Leanhardt, T.A. Pasquini, M.~Saba, A.~Schirotzek, Y.~Shin, D.~Kielpinski,
  D.E. Pritchard, and W.~Ketterle.
\newblock ``Cooling {B}ose-{E}instein condensates below 500 picokelvin''.
\newblock \href{https://dx.doi.org/10.1126/science.1088827}{Science {\bf 301},
  1513--1515}~(2003).

\bibitem{Olf2015}
Ryan Olf, Fang Fang, G.~Edward Marti, Andrew MacRae, and Dan~M Stamper-Kurn.
\newblock ``Thermometry and cooling of a {B}ose gas to 0.02 times the
  condensation temperature''.
\newblock \href{https://dx.doi.org/10.1038/nphys3408}{Nature Physics {\bf 11},
  720--723}~(2015).
\newblock  \href{http://arxiv.org/abs/1505.06196}{arXiv:1505.06196}.

\bibitem{Paris2015}
Matteo~G.A. Paris.
\newblock ``Achieving the {L}andau bound to precision of quantum thermometry in
  systems with vanishing gap''.
\newblock \href{https://dx.doi.org/10.1088/1751-8113/49/3/03lt02}{Journal of
  Physics A: Mathematical and Theoretical {\bf 49}, 03LT02}~(2015).
\newblock  \href{http://arxiv.org/abs/1510.08111}{arXiv:1510.08111}.

\bibitem{Mehboudi2019_1}
Mohammad Mehboudi, Anna Sanpera, and Luis~A Correa.
\newblock ``Thermometry in the quantum regime: recent theoretical progress''.
\newblock \href{https://dx.doi.org/10.1088/1751-8121/ab2828}{Journal of Physics
  A: Mathematical and Theoretical {\bf 52}, 303001}~(2019).
\newblock  \href{http://arxiv.org/abs/1811.03988}{arXiv:1811.03988}.

\bibitem{Cramer1999}
Harald Cram\'er.
\newblock ``Mathematical {M}ethods of {S}tatistics''.
\newblock Princeton University Press. ~(1999).
\newblock
  url:~\href{https://www.jstor.org/stable/j.ctt1bpm9r4}{www.jstor.org/stable/j.ctt1bpm9r4}.

\bibitem{Sondhi1997}
S.~L. Sondhi, S.~M. Girvin, J.~P. Carini, and D.~Shahar.
\newblock ``Continuous quantum phase transitions''.
\newblock \href{https://dx.doi.org/10.1103/revmodphys.69.315}{Reviews of Modern
  Physics\,\,{\bf 69}}~(1997).

\bibitem{Pelissetto2002}
Andrea Pelissetto and Ettore Vicari.
\newblock ``Critical phenomena and renormalization-group theory''.
\newblock \href{https://dx.doi.org/10.1016/s0370-1573(02)00219-3}{Physics
  Reports {\bf 368}, 549–727}~(2002).
\newblock
  \href{http://arxiv.org/abs/cond-mat/0012164}{arXiv:cond-mat/0012164}.

\bibitem{Fisher1972}
Michael~E. Fisher and Michael~N. Barber.
\newblock ``Scaling {T}heory for {F}inite-{S}ize {E}ffects in the {C}ritical
  {R}egion''.
\newblock \href{https://dx.doi.org/10.1103/PhysRevLett.28.1516}{Physical Review
  Letters {\bf 28}, 1516--1519}~(1972).

\bibitem{Botet1983}
R.~Botet and R.~Jullien.
\newblock ``Large-size critical behavior of infinitely coordinated systems''.
\newblock \href{https://dx.doi.org/10.1103/PhysRevB.28.3955}{Physical Review B
  {\bf 28}, 3955--3967}~(1983).

\bibitem{Rossini2018}
Davide Rossini and Ettore Vicari.
\newblock ``Ground-state fidelity at first-order quantum transitions''.
\newblock \href{https://dx.doi.org/10.1103/PhysRevE.98.062137}{Physical Review
  E\,\,{\bf 98}}~(2018).
\newblock  \href{http://arxiv.org/abs/1807.01674}{arXiv:1807.01674}.

\bibitem{Lacki2017}
Mateusz {\L}{\k{a}}cki and Bogdan Damski.
\newblock ``Spatial {K}ibble{\textendash}{Z}urek mechanism through
  susceptibilities: the inhomogeneous quantum {I}sing model case''.
\newblock \href{https://dx.doi.org/10.1088/1742-5468/aa8c20}{Journal of
  Statistical Mechanics: Theory and Experiment {\bf 2017}, 103105}~(2017).
\newblock  \href{http://arxiv.org/abs/1707.09884}{arXiv:1707.09884}.

\bibitem{Correa2015}
Luis~A. Correa, Mohammad Mehboudi, Gerardo Adesso, and Anna Sanpera.
\newblock ``Individual {Q}uantum {P}robes for {O}ptimal {T}hermometry''.
\newblock \href{https://dx.doi.org/10.1103/PhysRevLett.114.220405}{Physical
  Review Letters {\bf 114}, 220405}~(2015).
\newblock  \href{http://arxiv.org/abs/1411.2437}{arXiv:1411.2437}.

\bibitem{Lipkin1965}
H.J. Lipkin, N.~Meshkov, and A.J. Glick.
\newblock ``Validity of many-body approximation methods for a solvable model:
  (i). {E}xact solutions and perturbation theory''.
\newblock \href{https://dx.doi.org/10.1016/0029-5582(65)90862-X}{Nuclear
  Physics {\bf 62}, 188--198}~(1965).

\bibitem{Kawaguchi2012}
Yuki Kawaguchi and Masahito Ueda.
\newblock ``Spinor {B}ose--{E}instein condensates''.
\newblock \href{https://dx.doi.org/10.1016/j.physrep.2012.07.005}{Physics
  Reports {\bf 520}, 253 -- 381}~(2012).
\newblock  \href{http://arxiv.org/abs/1001.2072}{arXiv:1001.2072}.

\bibitem{Stamper-Kurn2013}
Dan~M. Stamper-Kurn and Masahito Ueda.
\newblock ``Spinor {B}ose gases: {S}ymmetries, magnetism, and quantum
  dynamics''.
\newblock \href{https://dx.doi.org/10.1103/RevModPhys.85.1191}{Rev. Mod. Phys.
  {\bf 85}, 1191--1244}~(2013).
\newblock  \href{http://arxiv.org/abs/1205.1888}{arXiv:1205.1888}.

\bibitem{Orenes2019}
Daniel~Benedicto Orenes, Anna~U Kowalczyk, Emilia Witkowska, and Giovanni
  Barontini.
\newblock ``Exploring the thermodynamics of spin-1 bose gases with synthetic
  magnetization''.
\newblock \href{https://dx.doi.org/10.1088/1367-2630/ab14b4}{New Journal of
  Physics {\bf 21}, 043024}~(2019).
\newblock  \href{http://arxiv.org/abs/1901.00427}{arXiv:1901.00427}.

\bibitem{Xue2018}
Ming Xue, Shuai Yin, and Li~You.
\newblock ``Universal driven critical dynamics across a quantum phase
  transition in ferromagnetic spinor atomic {B}ose-{E}instein condensates''.
\newblock \href{https://dx.doi.org/10.1103/PhysRevA.98.013619}{Physical Review
  A {\bf 98}, 013619}~(2018).
\newblock  \href{http://arxiv.org/abs/1805.02174}{arXiv:1805.02174}.

\bibitem{Dusuel2004}
S\'ebastien Dusuel and Julien Vidal.
\newblock ``Finite-{S}ize {S}caling {E}xponents of the
  {L}ipkin-{M}eshkov-{G}lick model''.
\newblock \href{https://dx.doi.org/10.1103/PhysRevLett.93.237204}{Physical
  Review Letters {\bf 93}, 237204}~(2004).

\bibitem{Evrard2020}
Bertrand Evrard, An~Qu, Jean Dalibard, and Fabrice Gerbier.
\newblock ``Production and characterization of a fragmented spinor
  {B}ose-{E}instein condensate''~(2020).
\newblock  \href{http://arxiv.org/abs/2010.15739}{arXiv:2010.15739}.

\bibitem{Langari2004}
A.~Langari.
\newblock ``Quantum renormalization group of {XYZ} model in a transverse
  magnetic field''.
\newblock \href{https://dx.doi.org/10.1103/physrevb.69.100402}{Physical Review
  B\,\,{\bf 69}}~(2004).

\bibitem{Franchini2017}
Fabio Franchini.
\newblock ``An {I}ntroduction to {I}ntegrable {T}echniques for
  {O}ne-{D}imensional {Q}uantum {S}ystems''.
\newblock \href{https://dx.doi.org/10.1007/978-3-319-48487-7}{Springer
  International Publishing}. ~(2017).
\newblock  \href{http://arxiv.org/abs/1609.02100}{arXiv:1609.02100}.

\bibitem{Affleck1999}
Ian Affleck and Masaki Oshikawa.
\newblock ``Field-induced gap in {C}u benzoate and other $s=\frac{1}{2}$
  antiferromagnetic chains''.
\newblock \href{https://dx.doi.org/10.1103/PhysRevB.60.1038}{Physical Review B
  {\bf 60}, 1038--1056}~(1999).
\newblock
  \href{http://arxiv.org/abs/cond-mat/9905002}{arXiv:cond-mat/9905002}.

\bibitem{Mikeska2004}
Hans-J{\"u}rgen Mikeska and Alexei~K. Kolezhuk.
\newblock ``One-dimensional magnetism''.
\newblock \href{https://dx.doi.org/10.1007/BFb0119591}{Chapter~1, pages 1--83}.
\newblock Springer Berlin Heidelberg. Berlin, Heidelberg~(2004).

\bibitem{Mehboudi2015}
Mohammad Mehboudi, Maria Moreno-Cardoner, Gabriele De~Chiara, and Anna Sanpera.
\newblock ``Thermometry precision in strongly correlated ultracold lattice
  gases''.
\newblock \href{https://dx.doi.org/10.1088/1367-2630/17/5/055020}{New Journal
  of Physics {\bf 17}, 055020}~(2015).
\newblock  \href{http://arxiv.org/abs/1501.03095}{arXiv:1501.03095}.

\bibitem{Hartmann2004_1}
Michael Hartmann, G\"unter Mahler, and Ortwin Hess.
\newblock ``Local versus global thermal states: Correlations and the existence
  of local temperatures''.
\newblock \href{https://dx.doi.org/10.1103/PhysRevE.70.066148}{Phys. Rev. E
  {\bf 70}, 066148}~(2004).
\newblock
  \href{http://arxiv.org/abs/quant-ph/0404164}{arXiv:quant-ph/0404164}.

\bibitem{Hartmann2004_2}
Michael Hartmann, G\"unter Mahler, and Ortwin Hess.
\newblock ``Existence of {T}emperature on the {N}anoscale''.
\newblock \href{https://dx.doi.org/10.1103/PhysRevLett.93.080402}{Phys. Rev.
  Lett. {\bf 93}, 080402}~(2004).
\newblock
  \href{http://arxiv.org/abs/quant-ph/0312214}{arXiv:quant-ph/0312214}.

\bibitem{Garcia-Saez2009}
Artur Garc\'{\i}a-Saez, Alessandro Ferraro, and Antonio Ac\'{\i}n.
\newblock ``Local temperature in quantum thermal states''.
\newblock \href{https://dx.doi.org/10.1103/PhysRevA.79.052340}{Phys. Rev. A
  {\bf 79}, 052340}~(2009).
\newblock  \href{http://arxiv.org/abs/0808.0102}{arXiv:0808.0102}.

\bibitem{Ferraro2012}
Alessandro Ferraro, Artur Garc{\'{i}}a-Saez, and Antonio Ac{\'{i}}n.
\newblock ``Intensive temperature and quantum correlations for refined quantum
  measurements''.
\newblock \href{https://dx.doi.org/10.1209/0295-5075/98/10009}{{EPL}
  (Europhysics Letters) {\bf 98}, 10009}~(2012).
\newblock  \href{http://arxiv.org/abs/1102.5710}{arXiv:1102.5710}.

\bibitem{Kliesch2019}
M.~Kliesch, C.~Gogolin, M.~J. Kastoryano, A.~Riera, and J.~Eisert.
\newblock ``Locality of {T}emperature''.
\newblock \href{https://dx.doi.org/10.1103/PhysRevX.4.031019}{Phys. Rev. X {\bf
  4}, 031019}~(2014).
\newblock  \href{http://arxiv.org/abs/1309.0816}{arXiv:1309.0816}.

\bibitem{Hernandez-Santana2015}
Senaida Hern{\'{a}}ndez-Santana, Arnau Riera, Karen~V. Hovhannisyan,
  Mart{\'{i}} Perarnau-Llobet, Luca Tagliacozzo, and Antonio Ac{\'{i}}n.
\newblock ``Locality of temperature in spin chains''.
\newblock \href{https://dx.doi.org/10.1088/1367-2630/17/8/085007}{New Journal
  of Physics {\bf 17}, 085007}~(2015).
\newblock  \href{http://arxiv.org/abs/1506.04060}{arXiv:1506.04060}.

\bibitem{Hernandez-Santana2021}
Senaida Hern{\'{a}}ndez-Santana, Andr{\'{a}}s Moln{\'{a}}r, Christian Gogolin,
  J.~Ignacio Cirac, and Antonio Ac{\'{i}}n.
\newblock ``Locality of temperature and correlations in the presence of
  non-zero-temperature phase transitions''.
\newblock \href{https://dx.doi.org/10.1088/1367-2630/ac14a9}{New Journal of
  Physics {\bf 23}, 073052}~(2021).
\newblock  \href{http://arxiv.org/abs/2010.15256}{arXiv:2010.15256}.

\bibitem{Palacios2018}
Silvana Palacios, Simon Coop, Pau Gomez, Thomas Vanderbruggen, Y.~Natali
  Martinez~de Escobar, Martijn Jasperse, and Morgan~W. Mitchell.
\newblock ``Multi-second magnetic coherence in a single domain spinor
  {B}ose{\textendash}{E}instein condensate''.
\newblock \href{https://dx.doi.org/10.1088/1367-2630/aab2a0}{New Journal of
  Physics {\bf 20}, 053008}~(2018).
\newblock  \href{http://arxiv.org/abs/1707.09607}{arXiv:1707.09607}.

\bibitem{Gomez2020}
Pau Gomez, Ferran Martin, Chiara Mazzinghi, Daniel Benedicto~Orenes, Silvana
  Palacios, and Morgan~W. Mitchell.
\newblock ``{B}ose-{E}instein {C}ondensate {C}omagnetometer''.
\newblock \href{https://dx.doi.org/10.1103/PhysRevLett.124.170401}{Physical
  Review Letters {\bf 124}, 170401}~(2020).
\newblock  \href{http://arxiv.org/abs/1910.06642}{arXiv:1910.06642}.

\bibitem{Eckert2008}
Kai Eckert, Oriol Romero-Isart, Mirta Rodriguez, Maciej Lewenstein, Eugene~S
  Polzik, and Anna Sanpera.
\newblock ``Quantum non-demolition detection of strongly correlated systems''.
\newblock \href{https://dx.doi.org/10.1038/nphys776}{Nature Physics {\bf 4},
  50--54}~(2008).
\newblock  \href{http://arxiv.org/abs/0709.0527}{arXiv:0709.0527}.

\bibitem{Len2021}
Yink~Loong Len, Tuvia Gefen, Alex Retzker, and Jan Ko{\l}ody{\'n}ski.
\newblock ``Quantum metrology with imperfect measurements''~(2021).
\newblock  \href{http://arxiv.org/abs/2109.01160}{arXiv:2109.01160}.

\bibitem{Plodzien2018}
Marcin P\l{}odzie\'{n}, Rafa\l{} Demkowicz-Dobrza\'{n}ki, and Tomasz
  Sowi\'{n}ski.
\newblock ``Few-fermion thermometry''.
\newblock \href{https://dx.doi.org/10.1103/PhysRevA.97.063619}{Physical Review
  A {\bf 97}, 063619}~(2018).
\newblock  \href{http://arxiv.org/abs/1804.04506}{arXiv:1804.04506}.

\end{thebibliography}

\end{document}